\documentclass[fleqn,usenatbib]{mnras}

\usepackage{newtxtext,newtxmath}
\usepackage{epsfig}
\usepackage{latexsym}
\usepackage{amsfonts}
\usepackage{bm}
\usepackage{multicol}

\usepackage[T1]{fontenc}

\DeclareRobustCommand{\VAN}[3]{#2}
\let\VANthebibliography\thebibliography
\def\thebibliography{\DeclareRobustCommand{\VAN}[3]{##3}\VANthebibliography}

\usepackage{graphicx}	
\usepackage{amsmath}	

\title[Trajectories and Radiation of Charged Particles in the Pulsar Magnetosphere]{Trajectories and Radiation of Charged Particles in the Pulsar Magnetosphere}

\author[S. Chang et al.]{
Shan Chang, Li Zhang \thanks{E-mail: lizhang@ynu.edu.cn}, Zejun Jiang, and Xiang, Li
\\
Department of Astronomy, Key Laboratory of Astroparticle Physics of Yunnan Province, Yunnan University, Kunming 650091
}

\date{Accepted XXX. Received YYY; in original form ZZZ}

\pubyear{2022}

\begin{document}
\label{firstpage}
\pagerange{\pageref{firstpage}--\pageref{lastpage}}
\maketitle

\begin{abstract}
Trajectories and radiation of the accelerating electrons are studied in the pulsar magnetosphere approximated as the electromagnetic field of the Deutsch's solutions. Because the electrons are accelerated rapidly to ultra-relativistic velocity near the neutron star surface, the electron velocity vector (and then its trajectory) is derived from the balance between Lorentz force and radiation reaction force, which makes the pitch angle between electron trajectories and magnetic field lines nonzero in most part of the magnetosphere. In such a case, the spectral energy distributions (SEDs) of synchro-curvature radiation for the accelerating electrons with a mono-energetic form are calculated. Our results indicate that: (i) the pitch angle is the function of electron position ($r, \theta, \phi$) in the open field line regions, and increases with increasing $r$ and $\theta$ as well as increasing the inclination angle; (ii) the radius of curvature becomes large along the particle trajectory, and (iii) the SED appears a double peak structure depending on the emission position, where the synchrotron radiation plays an important role in X-ray band and curvature radiation mainly works in GeV band, which is only determined by parameters $\alpha$ and $\zeta$.
\end{abstract}




\section{Introduction}

A pulsar magnetosphere is believed to be the site of acceleration and radiation of charged particles. However, the dynamics of particle acceleration and radiation mechanisms are not still clear. Up to now, more than 200 pulsars have been detected to emit pulsed $\gamma$-rays
\citep[e.g., ][]{Adbo2009,Adbo2010,Adbo2013,Ajello2017}. Based on the observations, the models for gamma-ray pulsars have been tested and/or developed, such as polar gap \citep[e.g., ][]{RS75,DH94,DH96}, slot gap \citep[e.g., ][]{Arons83,MH03,MH04}, outer gap \citep[e.g., ][] {Cheng1986,Zhang1997,Zhang2007,Chang2015}, annular gap \citep[e.g., ][]{Qiao2004,Du2011,Du2012}, and the striped wind \citep[e.g.,][]{Coroniti1990,PK2005,Petri2009,Chang2019a,Chang2019b} models. In these models, two basic assumptions are made:  the field structure inside the pulsar magnetosphere is approximated as a retarded dipole field, and the accelerated particles move outwards from the polar cap along the magnetic field lines because of strong magnetic field strength. Therefore, the curvature radiation is a dominating radiation mechanism if high-energy radiation occurs inside the light cylinder of the magnetosphere.

Although the gamma-ray pulsars models mentioned above have made some successful for explaining pulsed gamma-ray features \citep[e.g., ][]{CRZ2000,Dyks2003,Harding2008,Chang2018}, the above assumptions are apparently oversimplified. In fact, recently, great advances in numerical simulation of pulsar magnetospheres produce more realistic magnetospheres, and a particle trajectory approach is used to describe the particle trajectory in the magnetosphere \citep[e.g.,][]{KHKC12,KHK14,PSC15,CPS2016,KBTHK18}. Based on the simulations, the model which has
force-free inside and dissipative outside (FIDO) the light cylinder of the magnetosphere has been proposed to explain the features of gamma-ray pulsars \citep[e.g.,][]{KHK14,BKHK2015,KBTHK18}. Because of the deficiency of the numerical simulations, a hybrid approach has been proposed to simulate the pulsar magnetosphere \citep[e.g.,][]{Contopoulos2016,CPS2020}. On the other hand, almost vacuum electromagnetic fields can also be mimicked by the magnetosphere with low conductivity \citep[e.g.,][]{KHKC12,KHK14,GP2020}, and thus fields have analytic solutions.

In this paper, as an approximation, the pulsar magnetosphere inside the light cylinder is assumed to be described by the vacuum electromagnetic field given by \citet{Deutsch1955}. And then, the velocity of a particle that determines its trajectory in such a field is derived in detail after solving the kinematic equation of particles. Since the real particle trajectory is not always along the magnetic field line but with a pitch angle between the magnetic field line and trajectory, the properties of non-thermal photons produced by synchro-curvature radiation mechanism at different locations of the magnetosphere are studied in detail. In this paper, the parameters of the Crab pulsar are used, such as the period $P\approx 1/30$ s and the strength of magnetic field on the surface of the neutron star $B_{*}\approx 3.8\times 10^{12}$ G. The paper is organized as follows. In section \ref{sec:trjectory}, the electromagnetic field of the Deusch's solution is reviewed briefly, and the trajectories of the particles inside the magnetosphere are calculated. In section \ref{sec:Calculation-Procedure}, the calculation procedure is given, and results are given in section \ref{sec:Results}. In section \ref{sec:summary and discussion}, the results are discussed and our conclusions are given.

\section{Particle trajectories and radiations in electromagnetic fields}\label{sec:trjectory}

\subsection{The Deutsch's solution of electromagnetic fields}

As mentioned above, the pulsar magnetosphere is approximated as the vacuum electromagnetic field given by \citet{Deutsch1955}. Here, the recast version of Deutsch's solution given by \citet{Michel&Li1999} is used. In the spherical polar coordinate($r, \theta, \phi$), the magnetic field is written as,
\begin{eqnarray}\label{Eq1}
    \nonumber  && B_r= 2 B_{*}\frac{R_{*}^3}{r^3}\left\{\cos\alpha\cos\theta + \sin\alpha\sin\theta[d_1\cos\psi + d_2\sin\psi]\right\}   \;, \\
    \nonumber  && B_\theta=\ B_{*} \frac{R_{*}^3}{r^3}\left\{\cos\alpha\sin\theta  \right. \\
               && \left.\qquad -\sin\alpha\cos\theta[(q_1+d_3)\cos\psi+(q_2+d_4)\sin\psi]\right\}   \;, \\
    \nonumber  && B_\varphi=\ B_{*} \frac{R_{*}^3}{r^3}\sin\alpha\left\{-[q_2\cos2\theta+d_4]\cos\psi \right. \\
    \nonumber  && \left. \qquad +[q_1\cos2\theta+d_3]\sin\psi\right\}   \;,
\end{eqnarray}
and the electric field is given by,
\begin{eqnarray}\label{Eq2}
    \nonumber  && E_r=E_{*}\frac{R_{*}^2}{r^{2}}\left\{\frac{2}{3}\cos\alpha + \frac{R_{*}^2}{r^2}\cos\alpha (1-3\cos^{2}\theta) \right. \\
    \nonumber  && \left. \qquad -\frac{3}{\rho^2}\sin\alpha\sin 2\theta[q_1\cos\psi+q_2\sin\psi] \right\}   \;, \\
               && E_\theta=E_{*} \frac{R_{*}^2}{r^2}\left\{-\frac{R_{*}^{2}}{r^2}\cos\alpha\sin2\theta \right. \\
    \nonumber  && \left.\qquad +\sin\alpha[(q_3\cos2\theta-d_1)\cos\psi+(q_4\cos2\theta-d_2)\sin\psi] \right\}   \;, \\
    \nonumber  && E_\varphi=E_{*} \frac{R_{*}^2}{r^2}\sin\alpha\cos\theta\left\{(q_4-d_2)\cos\psi-(q_3-d_1)\sin\psi \right\} \;,
\end{eqnarray}
where $\psi=\phi_{\rm s}+\rho-a$ and $\phi_{\rm s}=\phi-\Omega t$, the extra term $\rho-a$ comes from the Bessel functions. $a$ and $\rho$ represent the distance in unit of the radius of the light cylinder $R_{\rm L}=c/\Omega$ with an angular speed $\Omega$ inside the light cylinder, $a=R_{*}/R_{\rm L}$ and $\rho=r/R_{\rm L}$, where $R_*$ is the radius of the neutron star and $r$ is the radial distance to the star. $B_*$ and $E_*$ are the strength of magnetic field and electric field on the surface of neutron star, which can be estimated by $B_*=3.2\times 10^{19} P \dot{P}\ \rm G$ and $E_*=\Omega R_* B_*/c$.

In above equations, the $d_i$ and $q_i$ terms are related to dipole and quadrupole. They can be expressed by $a$ and $\rho$ as follows \citep{Michel&Li1999},
\begin{eqnarray}\label{Eq3}
    \nonumber  &&  \rm d_1=\frac{a\rho+1}{a^2+1}  \;, \\
               &&  \rm d_2=\frac{\rho-a}{a^2+1}  \;,\\
    \nonumber  &&  \rm d_3=\frac{1+a\rho-\rho^2}{a^2+1}  \;, \\
    \nonumber  &&  \rm d_4=\frac{(\rho^2-1)a+\rho}{a^2+1}  \;,
\end{eqnarray}
and
\begin{eqnarray}\label{Eq4}
    \nonumber  &&  \rm q_1=\frac{3\rho(6a^3-a^5)+(3-\rho^2)(6a^2-3a^4)}{a^6-3a^4+36}  \;, \\
               &&  \rm q_2=\frac{(3-\rho^2)(a^5-6a^3)+3\rho(6a^2-3a^4)}{a^6-3a^4+36}  \;,\\
    \nonumber  &&  \rm q_3=\frac{(\rho^3-6\rho)(a^5-6a^3)+(6-3\rho^2)(6a^2-3a^4)}{\rho^2(a^6-3a^4+36)}  \;, \\
    \nonumber  &&  \rm q_4=\frac{(6-3\rho^2)(a^5-6a^3)+(6\rho-\rho^3)(6a^2-3a^4)}{\rho^2(a^6-3a^4+36)}  \;.
\end{eqnarray}
When $\rho \sim a$, then $d_1\approx 1$, $d_3 \approx 1$, $q_1 \approx a^2/2$, $q_3\approx(a/\rho)^2$ and $\psi=\phi_{\rm s}$, Eqs.(\ref{Eq1}) and (\ref{Eq2}) can be simplified (see Eq.(92) of \cite{Michel&Li1999}), that represent fields of near region. And when $\rho \gg 1$, then $d_2 \approx\rho$ and $d_3\approx-\rho^2$, equations represent fields of far region (also see Eq.(94) of \cite{Michel&Li1999}). Then the 3D structure of the pulsar magnetosphere can be simulated by using above equations, where the field lines are described by the reduced factor ranges from $0$ (magnetic axis) to $1$ (the last open field lines).

\subsection{Particle trajectories}

\begin{figure*}
      \centering
      \includegraphics[width=0.45\textwidth]{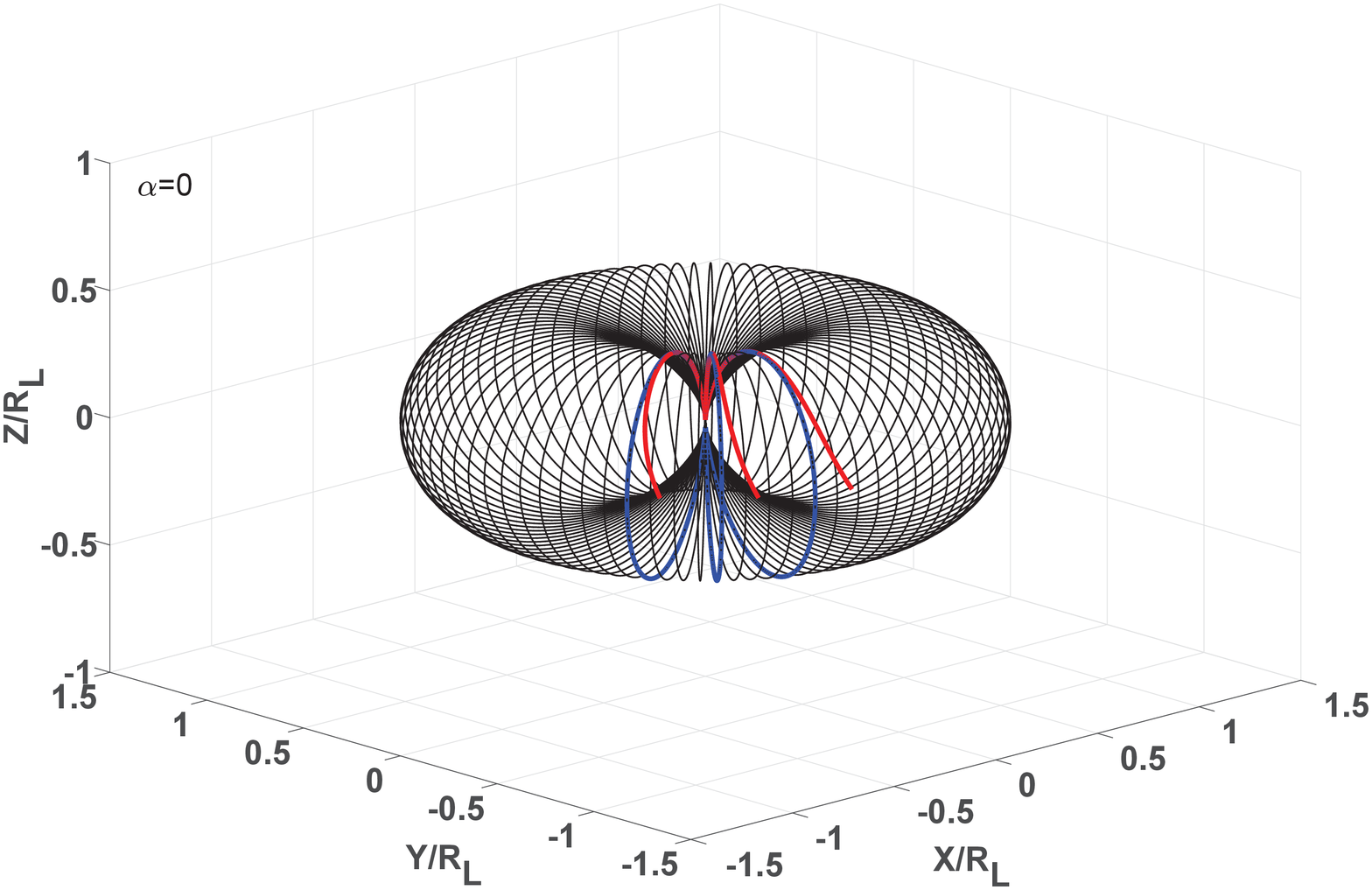}
      \includegraphics[width=0.45\textwidth]{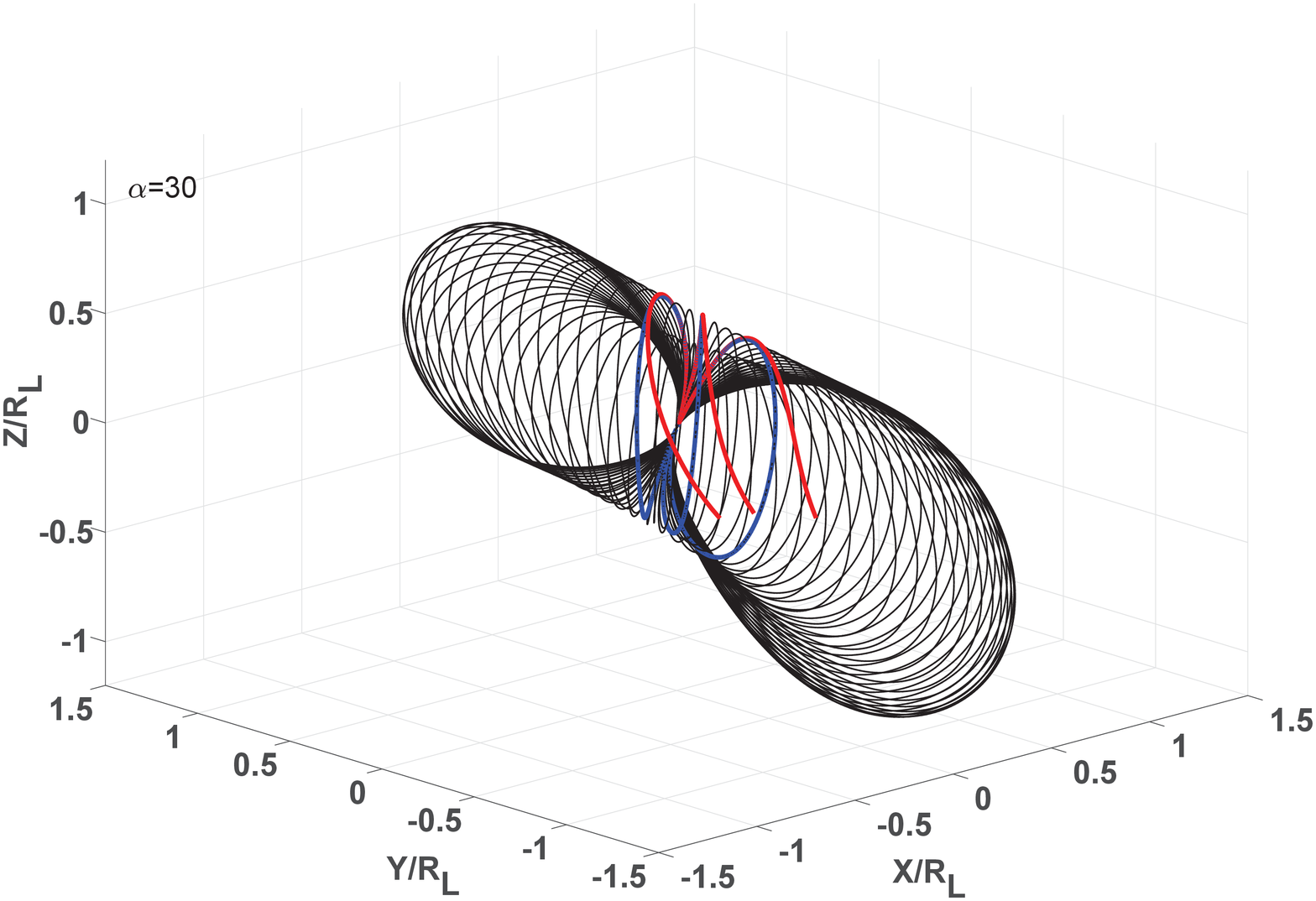}
      \includegraphics[width=0.45\textwidth]{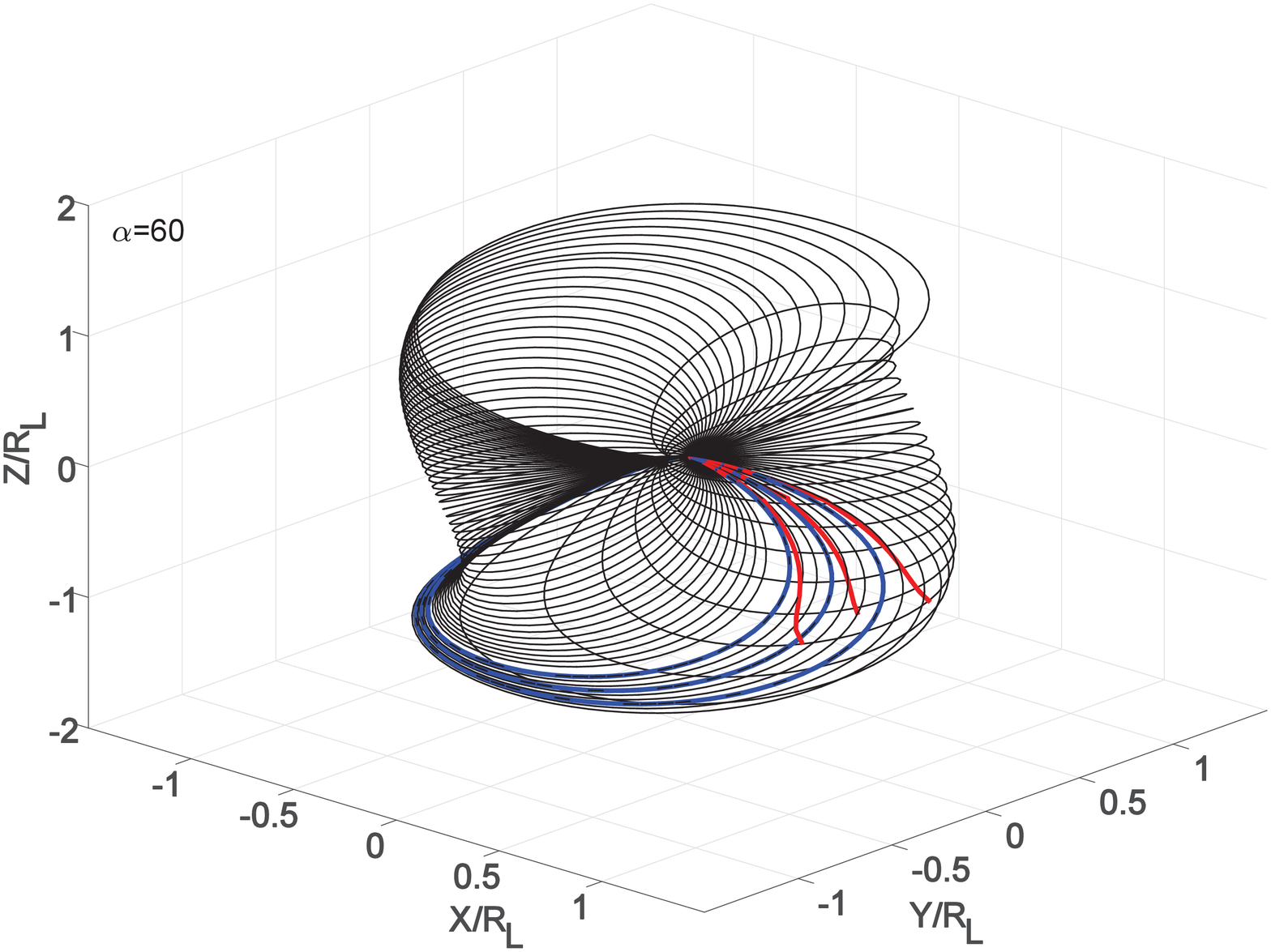}
      \includegraphics[width=0.45\textwidth]{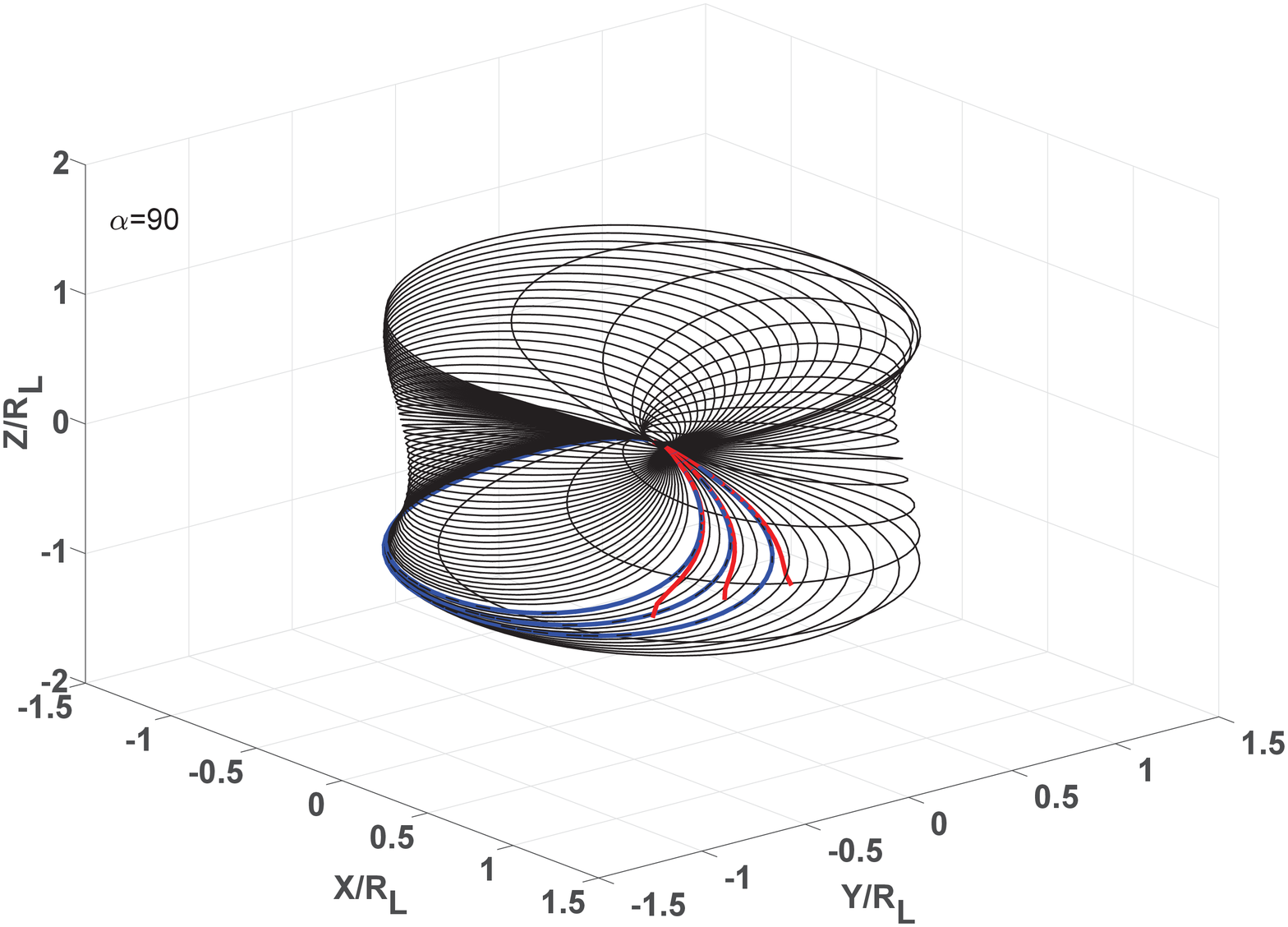}
      \caption{Particle trajectories compared to last closed field lines for four oblique rotators ($\alpha=0^{\circ}$, $30^{\circ}$, $60^{\circ}$, and $90^{\circ}$) under Deutsch's electromagnetic field in three-dimensional Cartesian Coordinates. Black and blue lines represent magnetic field lines, and red lines represent trajectories of electrons. }
      \label{fig:1}
\end{figure*}

\begin{figure*}
      \centering
      \includegraphics[width=0.42\textwidth]{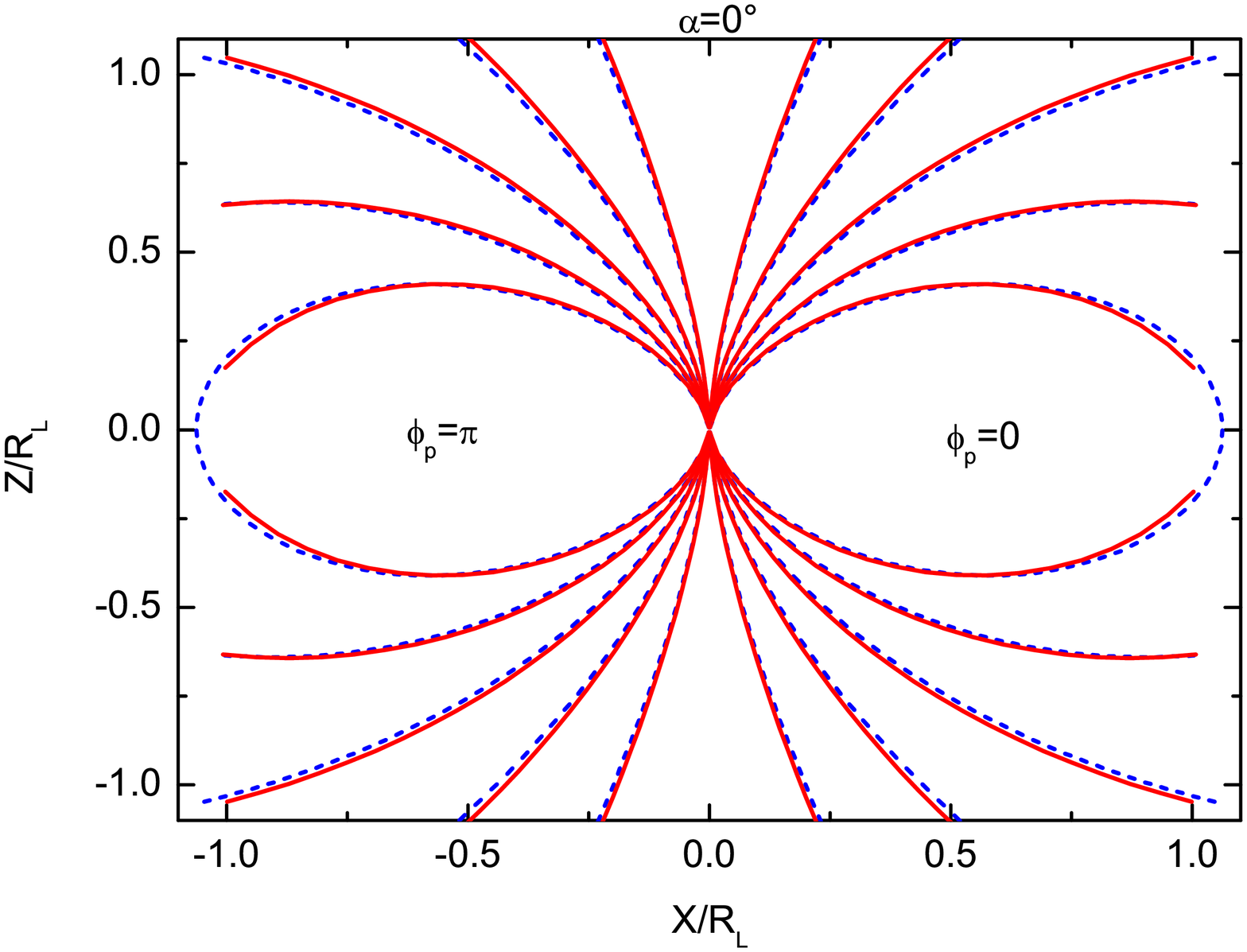}
      \includegraphics[width=0.42\textwidth]{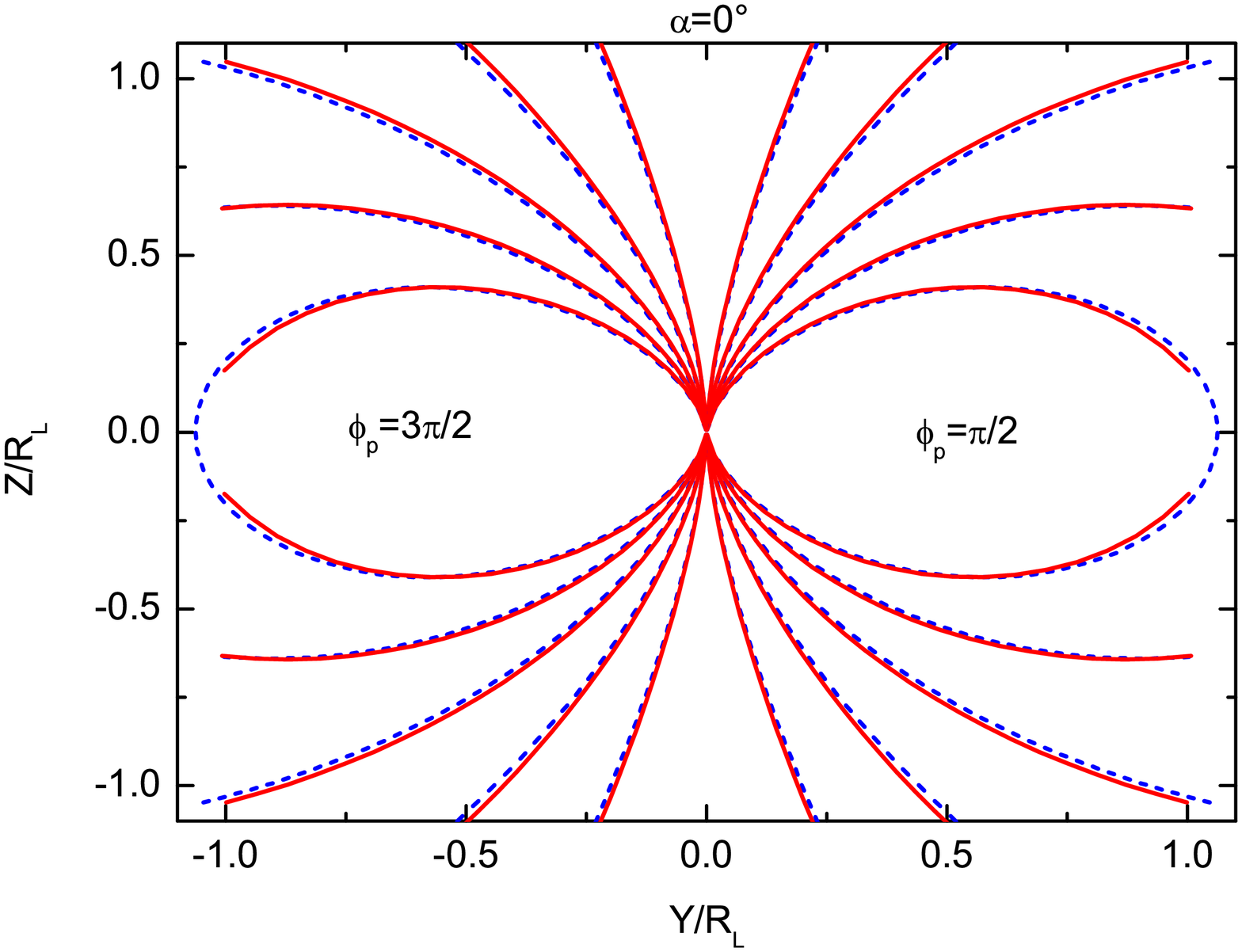}
      \includegraphics[width=0.42\textwidth]{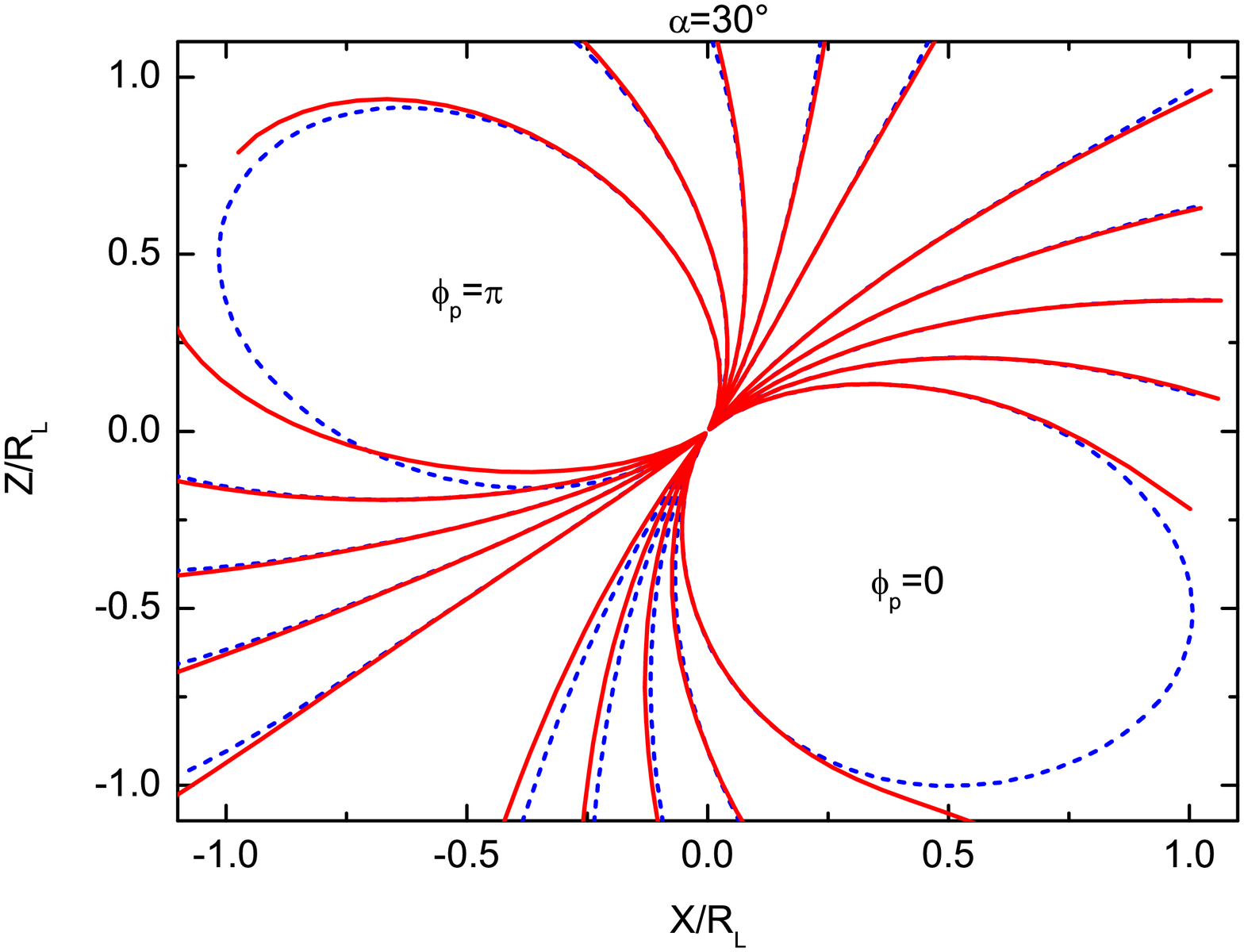}
      \includegraphics[width=0.42\textwidth]{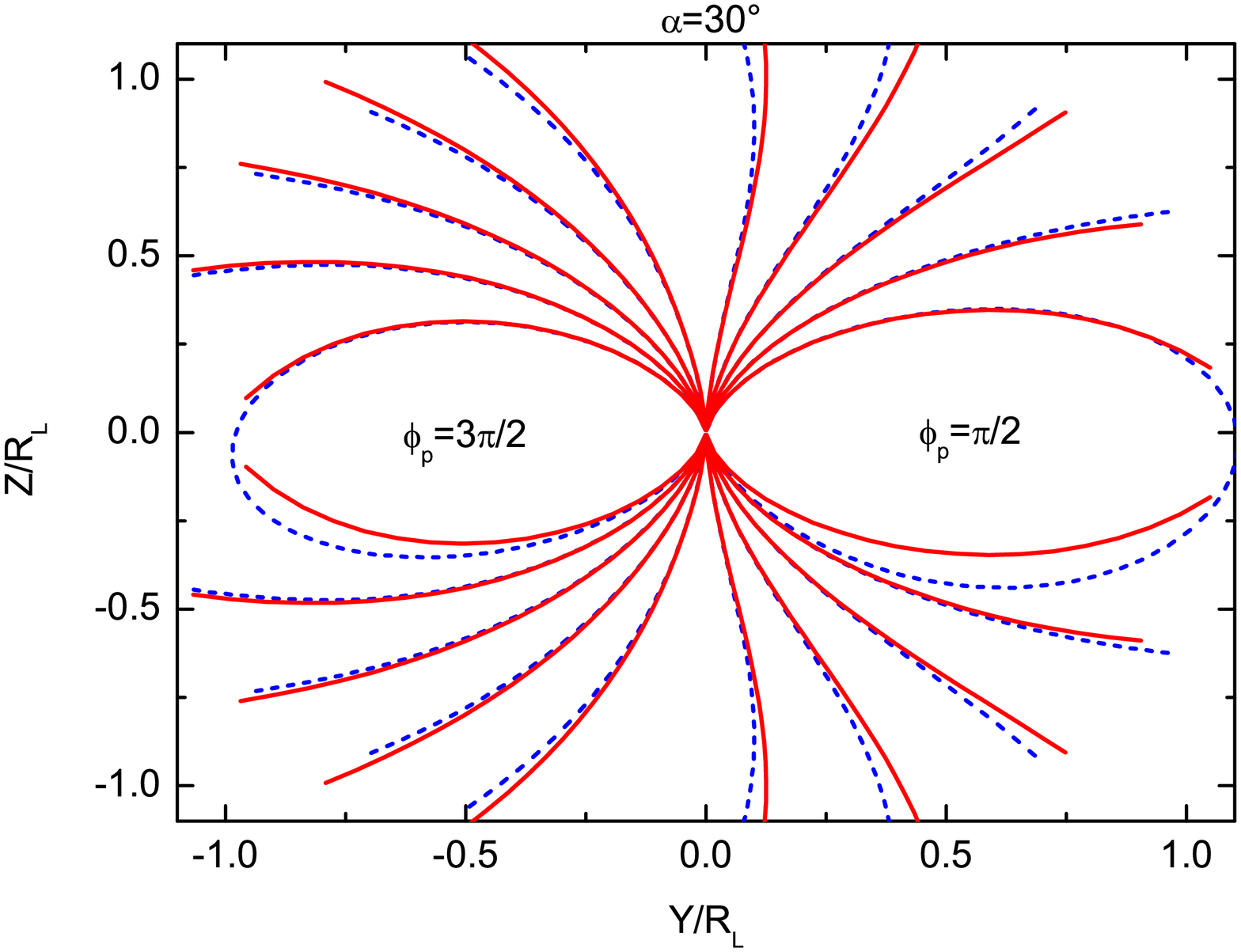}
      \includegraphics[width=0.42\textwidth]{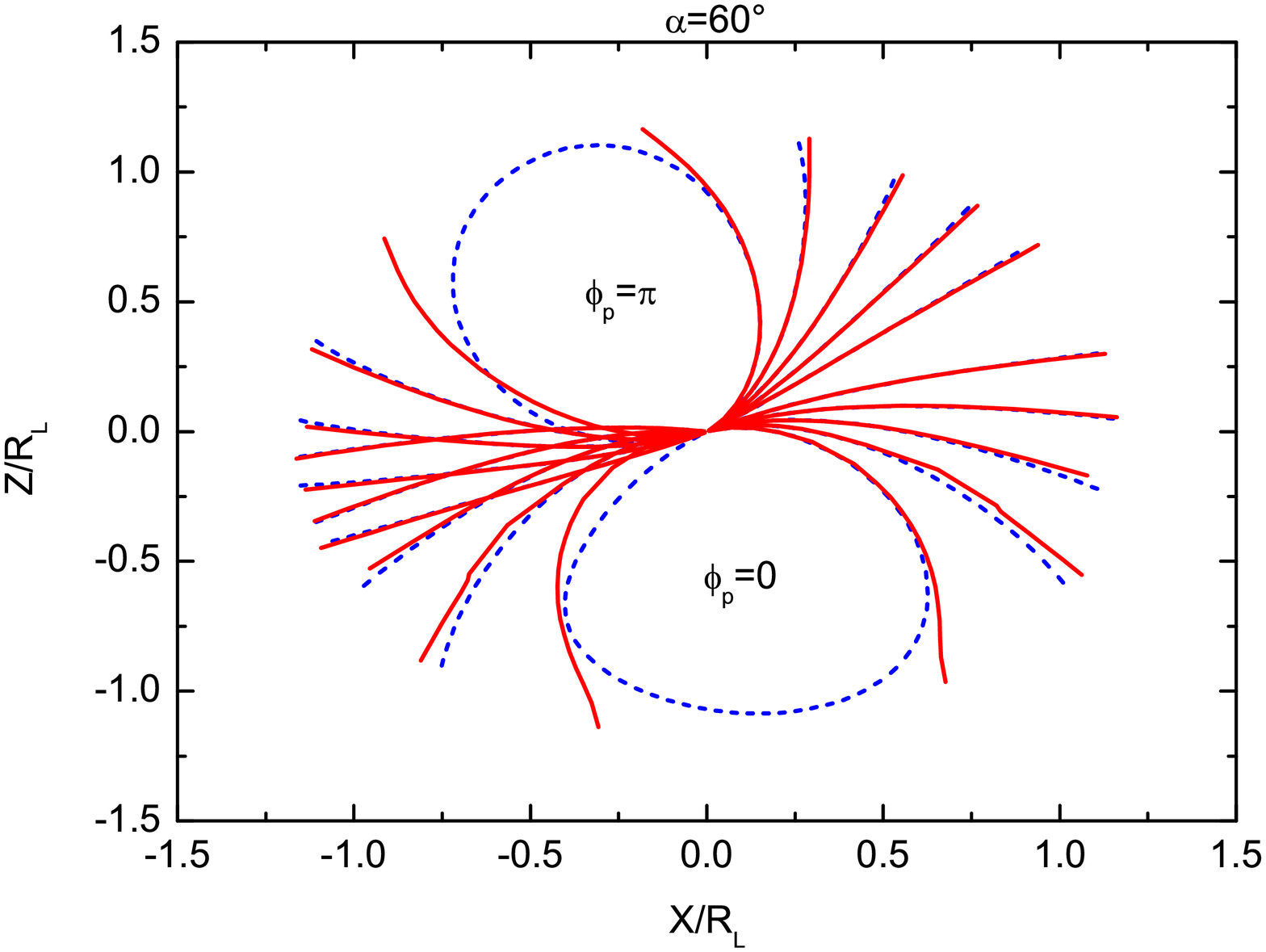}
      \includegraphics[width=0.42\textwidth]{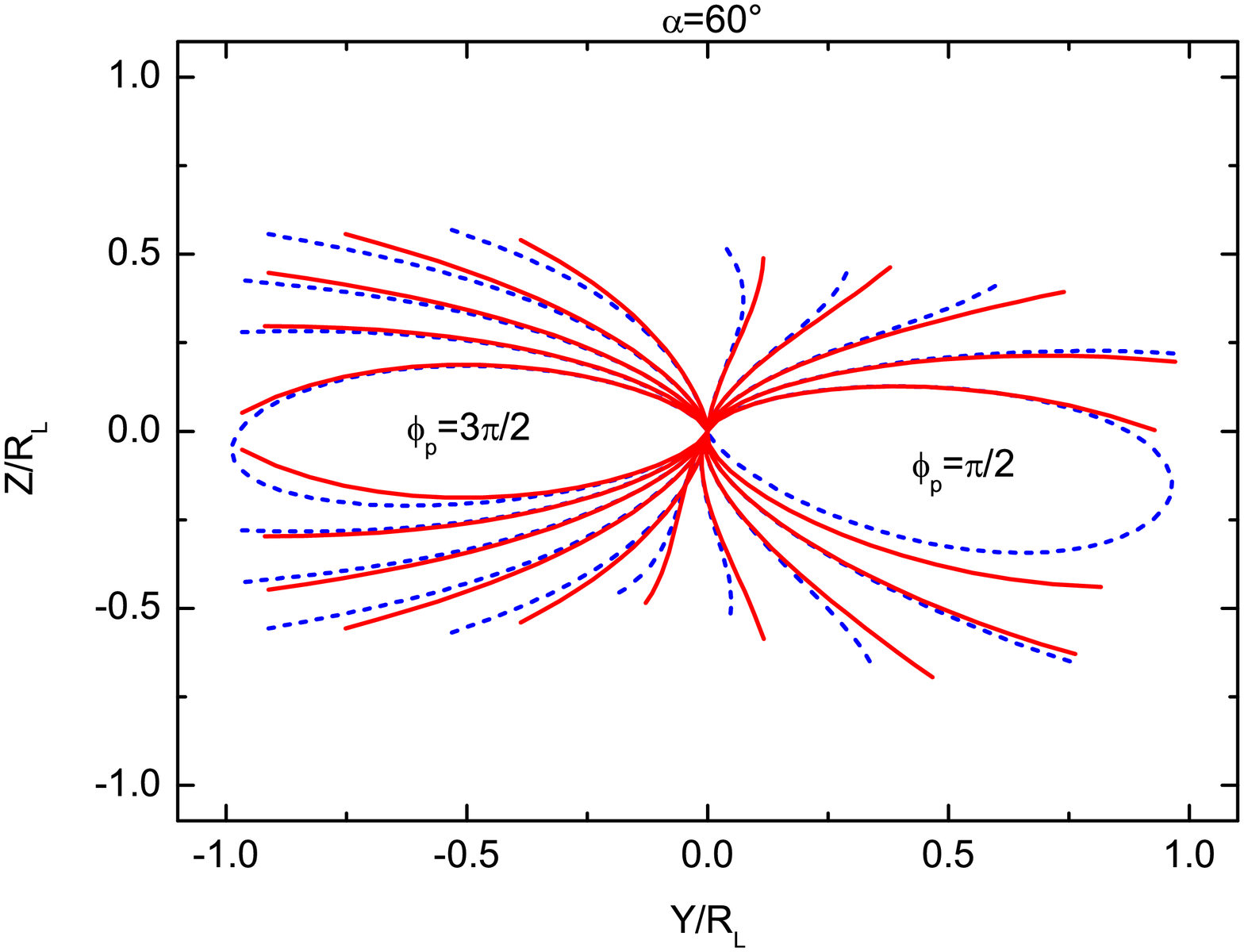}
      \includegraphics[width=0.42\textwidth]{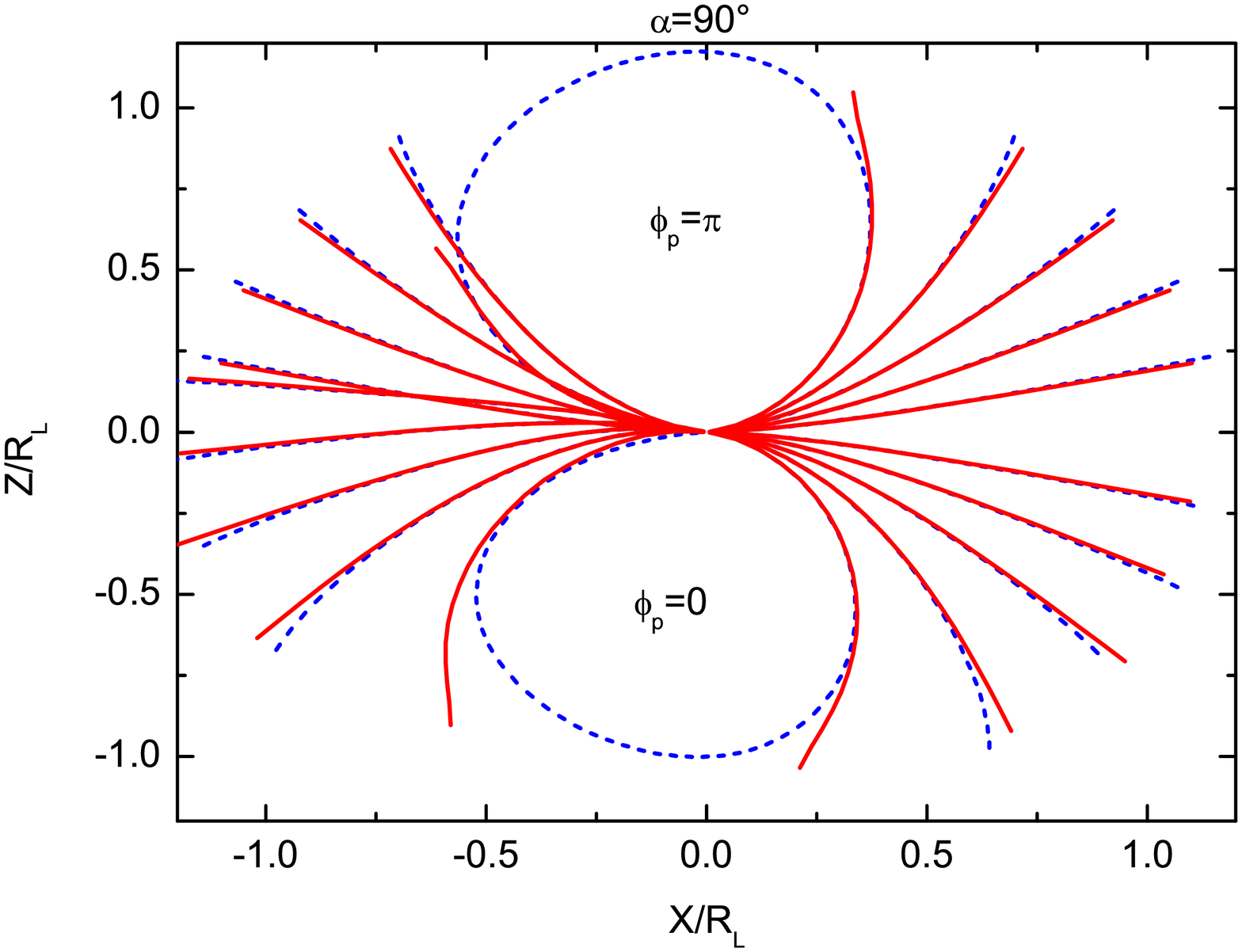}
      \includegraphics[width=0.4\textwidth]{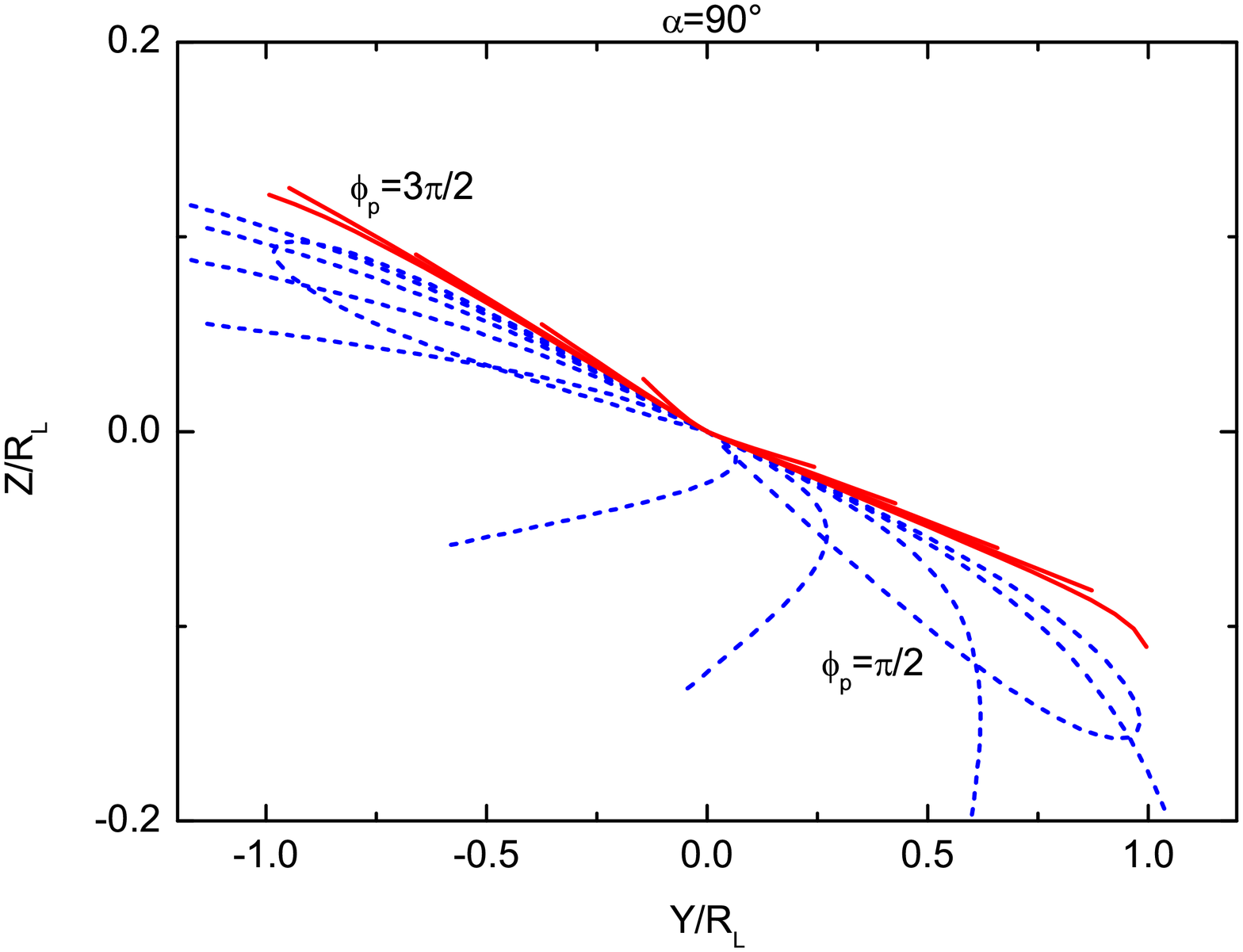}
      \caption{Comparison of open magnetic field lines and particle trajectories on X-Z and Y-Z planes for obliquities $\alpha\in\{ 0^{\circ}, 30^{\circ}, 60^{\circ}, 90^{\circ} \}$. Blue dash lines show open field lines ranged from $a_i=1,~0.8,~0.6,~0.4,~0.2$, and red solid lines show trajectories of electrons. Notes that the azimuthal angel $\phi_p=0$ and $\pi$ are shown on X-Z plane as well as $\phi_p=\pi/2$ and $3\pi/2$ are shown on Y-Z plane.}
      \label{fig:2}
\end{figure*}

The motion of an ultra-relativistic charged particle in the electric($\textbf{E}$) and magnetic($\textbf{B}$) fields can be expressed as,
\begin{equation}\label{Eq5}
  \frac{d\bm{p}}{dt}=\textbf{F}_{\rm L}-\textbf{F}_{\rm rr}  \;,
\end{equation}
where $\bm{p}$ is the momentum of the relativistic particle, $e$ is the electron charge, $\textbf{F}_{\rm L}$ is the Lorentz force, and $\textbf{F}_{\rm rr}$ represents the radiation reaction force \citep{Landau&Lifshitz1987}. The energy ($\varepsilon$) of relativistic particles is expressed as $\varepsilon=m_ec^2/\sqrt{1-\beta^2}$, where $\bm{\beta}=\bm{v}/c$ is the relative velocity of ultra-relativistic charged particle. Furthermore, from the mass-energy equivalence, the derivation of  the particle energy can be given by
$\varepsilon(d\varepsilon/dt)=[\bm{p}\cdot(d\bm{p}/dt)]c^2$. Since $\bm{p}=\bm{v}\cdot \varepsilon/c^2$ for relativistic particles, $d\varepsilon/dt=\bm{v}\cdot (d\bm{p}/dt)$, i.e.,
\begin{equation}\label{Eq6}
  \frac{d\varepsilon}{dt}=\bm{v}\cdot(\textbf{F}_{\rm L}-\textbf{F}_{\rm rr}) \;.
\end{equation}
In addition, the derivative of $\bm{p}=\bm{v}\cdot\varepsilon/c^2$ is $ d\bm{p}/dt=(\varepsilon/c)(d\bm{\beta}/dt)+(\bm{\beta}/c)(d\varepsilon/dt)$. Therefore, the derivative of $\bm{\beta}$ is given by
\begin{equation}\label{Eq7}
  \frac{d\bm{\beta}}{dt}=\frac{1}{m_ec\gamma_e}\left\{ (\textbf{F}_{\rm L}-\textbf{F}_{\rm rr})-\bm{\beta}[\bm{\beta}\cdot(\textbf{F}_{\rm L}-\textbf{F}_{\rm rr})] \right\}  \;,
\end{equation}
where $\varepsilon$ can also be represented by $\varepsilon=\gamma_em_ec^2$, and $\gamma_e$ is the Lorentz factor of the electron.

Thus, Eq.(\ref{Eq7}) becomes
\begin{equation}\label{Eq8}
  (\textbf{F}_{\rm L}-\textbf{F}_{\rm rr})-\bm{\beta}[\bm{\beta}\cdot(\textbf{F}_{\rm L}-\textbf{F}_{\rm rr})]- m_{\rm e}c\gamma_{\rm e} \frac{d\bm{\beta}}{dt}=0 \;,
\end{equation}
where the Lorentz force is $\textbf{F}_{\rm L}=e(\textbf{E}+\bm{\beta}\times\textbf{B})$, and the radiation reaction force is described as\citep{Landau&Lifshitz1987},
\begin{align}\label{Eq9}
  \nonumber \textbf{F}_{\rm rr} &=\frac{2e^4}{3m_{e}^{2}c^4}[\textbf{E}\times\textbf{B}+\textbf{B}\times(\textbf{B}\times\bm{\beta})+\textbf{E}(\bm{\beta}\cdot\textbf{E})] \\
                            &-\frac{2e^4\gamma_e^2}{3m_e^2c^4}\bm{\beta}[(\textbf{E}+\bm{\beta}\times\textbf{B})^2-(\textbf{E}\cdot\bm{\beta})^2] \\
  \nonumber                 &+\frac{2e^3\gamma_e^2}{3m_ec^3}[(\frac{\partial}{\partial t}+c\bm{\beta}\cdot\nabla)\textbf{E}+\bm{\beta}\times(\frac{\partial}{\partial t}+\bm{\beta}\cdot\nabla)\textbf{B}] \;,
\end{align}
the third term on the right hand of Eq.(\ref{Eq9}) contains the convection derivative of the field, which is difficult to implement. It can be found that it can be ignored compared to the first two terms.

Note that all the calculations in this paper are carried out in Gauss unit.

From Eq.(\ref{Eq8}), the particle velocity is given by (the derivation sees Appendix \ref{appendix A} in detail),
\begin{equation}\label{Eq10}
  \rm \bm{\beta}=\frac{(\textbf{E}\times\textbf{B})\pm (B_0\textbf{B}+E_0\textbf{E})}{B^2} \;,
\end{equation}
where $B_0$ and $E_0$ are given by Eq.(\ref{EqA10}), which have the same dimensions with $B$ and $E$. Two signs correspond to the two type of charges. Eq. (\ref{Eq10}) also can be written as
\begin{equation}\label{Eq11}
\bm{\beta}=\bm{\beta}_{\rm D}+\bm{\beta}_{\rm B}+\bm{\beta}_{\rm E}\;,
\end{equation}
i.e., the trajectories of particles are not strictly along the magnetic field lines, but consist of three components. On the right side of Eq.(\ref{Eq11}), the first term represents the drift velocity component, namely $\bm{\beta}_{\rm D}=(\textbf{E}\times\textbf{B})/B^2$; the second term is the velocity component along the magnetic field, namely $\bm{\beta}_{\rm B}=\pm B_0\bm{B}/B^2$; and the third term represents the velocity component along the electric field, namely $\bm{\beta}_{\rm E}=\pm E_0\bm{E}/B^2$. Therefore, the particle's velocity is the function of local ${\bf B}$ and ${\bf E}$ and determines the particle's trajectory.

Once the particle's trajectory is determined and the value of $\bm{\beta}$ is given, the particle's Lorentz factor is calculated by
\begin{equation}\label{Eq12}
\gamma_{\rm e} =\frac{1}{\sqrt{1 -|\bm{\beta}|^2}}\;.
\end{equation}
Note that $\gamma_{\rm e}$ changes with location of the particles. Moreover, the pitch angle between a trajectory and a magnetic field line is calculated by
\begin{equation}\label{Eq13}
\theta_0=\arccos(\beta_{\|}/\beta)\;.
\end{equation}
$\beta_{\|}$ is the velocity component parallel to the direction of the magnetic field.

\subsection{Synchro-curvature radiation}

To calculate the spectrum of the particle's radiation, concrete radiation mechanisms are required. Here, the synchro-curvature radiation is considered.  Following \cite{Cheng&Zhang1996} and \cite{Zhang1997}, the synchro-curvature spectrum of charged particles with relativistic energy of $\gamma_e$ can be written as,
\begin{equation}\label{Eq14}
   \frac{d^2N}{dE_{\gamma}dt}=\int_{E_{\rm{e,min}}}^{E_{\rm{e}}}\left (\frac{dN_{\rm e}}{dE'_{\rm e}}\right )\left (\frac{d^2 N}{dE_{\gamma}dt}\right ) dE'_{\rm e} \;,
\end{equation}
where $E_{\rm{e,min}}$ and $E_{\rm{e}}=m_ec^2\gamma_{\rm e}$ are the minimum and maximum energies of the particles, respectively, $dN_{\rm e}/dE'_{\rm e}$ is the electron spectrum distribution, and $d^2N/dE_\gamma dt$ represents the synchro-curvature spectrum of a single particle.

If the particles are injected on the stellar surface in a mono-energetic form, i.e. $dN_{\rm e}/dE'_{\rm e}\propto \delta(\gamma'_{\rm e} -\gamma_{\rm e})$. In this case, the synchro-curvature radiation can be described as that with a single Lorentz factor $\gamma_{\rm e, max}$, and the synchro-curvature spectrum of a single particle is expressed as
\begin{equation}\label{Eq15}
   \frac{d^2N}{dE_{\gamma}dt}=\left(\frac{dN^2}{dE_{\gamma}dt}\right)_{\|} + \left(\frac{d^2N}{dE_{\gamma}dt}\right)_{\perp}  \;,
\end{equation}
\begin{eqnarray}\label{Eq16}
   \nonumber   \left(\frac{d^2N}{dE_{\gamma}dt}\right)_{\|}=\frac{\sqrt{3}z e^2 \gamma_{\rm e}}{4\pi\hbar r_{\rm eff}E_{\gamma}}[F(y)+G(y)] \;, \\
               \left(\frac{d^2N}{dE_{\gamma}dt}\right)_{\perp}=\frac{\sqrt{3} e^2 \gamma_{\rm e}}{4\pi\hbar r_{\rm eff}E_{\gamma}}[F(y)-G(y)]\;,
  \end{eqnarray}
  $\hbar$ and $e$ are the reduced Planck constant and electron charge, respectively. $E_{\gamma}$ represent the energy of radiated photon.  To complete, other quantities in Eq.(\ref{Eq16}) are also listed as follows,
  \begin{eqnarray}\label{Eq17}
    \nonumber  && \Omega_0=\frac{c\cdot\cos\theta_0}{r_{\rm c}}  \;, \\
    \nonumber  && \omega_{\rm B}=\frac{eB}{\gamma_{\rm e}m_{\rm e}c}  \;,\\
               &&  r_{\rm B}=\frac{c\cdot\sin\theta_0}{\omega_{\rm B}}  \;, \\
    \nonumber  &&  r_{\rm eff}=\frac{c^2}{[(r_{\rm B}+r_{\rm c})\Omega_0^2+r_{\rm B}\omega_{\rm B}^2]}  \;,\\
    \nonumber  &&  Q_2^2=\frac{1}{r_{\rm B}}\times \;,\\
    \nonumber  &&  \left( \frac{r_{\rm B}^2+r_{\rm c}r_{\rm B}-3r_{\rm c}^2}{r_{\rm c}^3}\cos^4\theta_0+\frac{3}{r_{\rm c}}\cos^2\theta_0+\frac{1}{r_{\rm B}}\sin^4\theta_0 \right)  \;,\\
    \nonumber  &&  z=(Q_2r_{\rm eff})^{-2} \;, \\
    \nonumber  &&  F(y)=y\int_{y}^{\infty} K_{5/3}(x)dx \;,\\
    \nonumber  &&  G(y)=yK_{2/3}(y) \;,
\end{eqnarray}
where $\Omega_0$, $\omega_{\rm B}$ and $r_{\rm B}$ represent the angle velocity of the moving electron's guiding centre, Larmor frequency and Larmor radius of electron, respectively. $\theta_0$, $r_{\rm eff}$ and $r_{\rm c}$ are the pitch angle between trajectories and magnetic field lines, the instantaneous curvature of its trajectory, and the curvature radius of the trajectories, respectively. $K_{2/3}$ and $K_{5/3}$ are the second kind of modified Bessel functions of order $2/3$ and $5/3$, where $y=E_{\gamma}/E_{\rm c}^{\rm sc}$ and $E_{\rm c}^{\rm sc}=\frac{3}{2}\hbar c Q_2\gamma_{\rm e}^3$ is the characteristic energy of synchro-curvature radiation. Note that the point on the particle trajectory corresponds to the point on the magnetic field lines. Therefore, the real radii of curvature are consist of two parts: one part is the radius of curvature relative to the point on the magnetic field line, which is calculated by using Eq.(\ref{EqB5}), and another is the radius of curvature of the magnetic field line.

The power of the synchro-curvature radiation for a single electron is given by \citep{Cheng&Zhang1996,Zhang1997}
\begin{equation}\label{Eq18}
  P_{\rm syn-cur}=-\frac{e^2\gamma_{\rm e}^4cQ_2}{12r_{\rm eff}}\left(1+\frac{7}{r_{\rm eff}^2Q_{2}^2}\right)  \;,
\end{equation}
for the synchrotron radiation, we have
\begin{equation}\label{Eq19}
  P_{\rm syn}=-\frac{2e^4B^2\sin^2\theta_0\gamma_{\rm e}^2}{3m_{\rm e}^2c^3}  \;,
\end{equation}
and for the curvature radiation, we have
\begin{equation}\label{Eq20}
  P_{\rm cur}=-\frac{2e^2c\gamma_{\rm e}^4}{3r_{\rm c}^2}  \;.
\end{equation}

\section{Calculation Procedure}\label{sec:Calculation-Procedure}

\begin{figure*}
      \centering
      \includegraphics[width=0.48\textwidth]{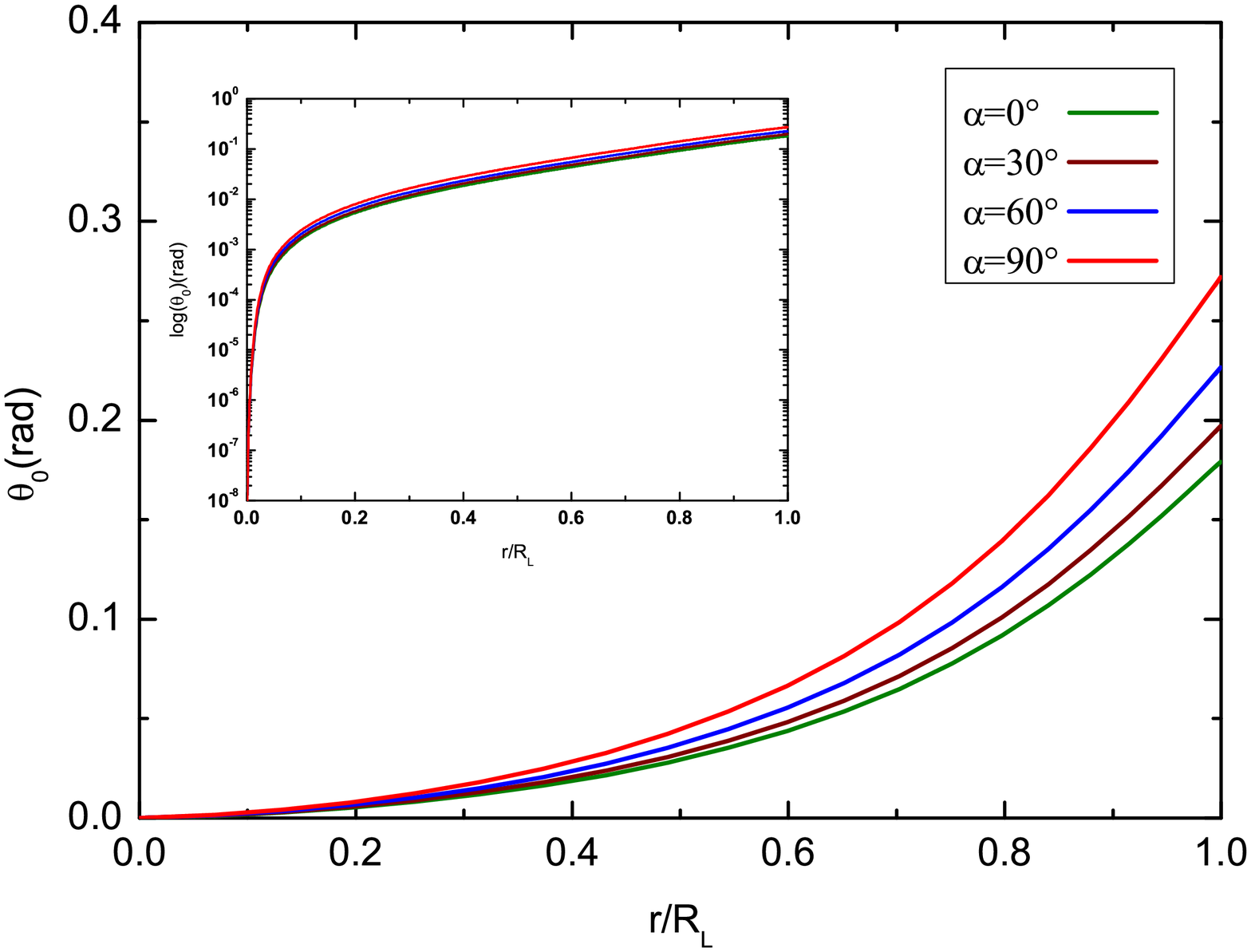}
      \includegraphics[width=0.48\textwidth]{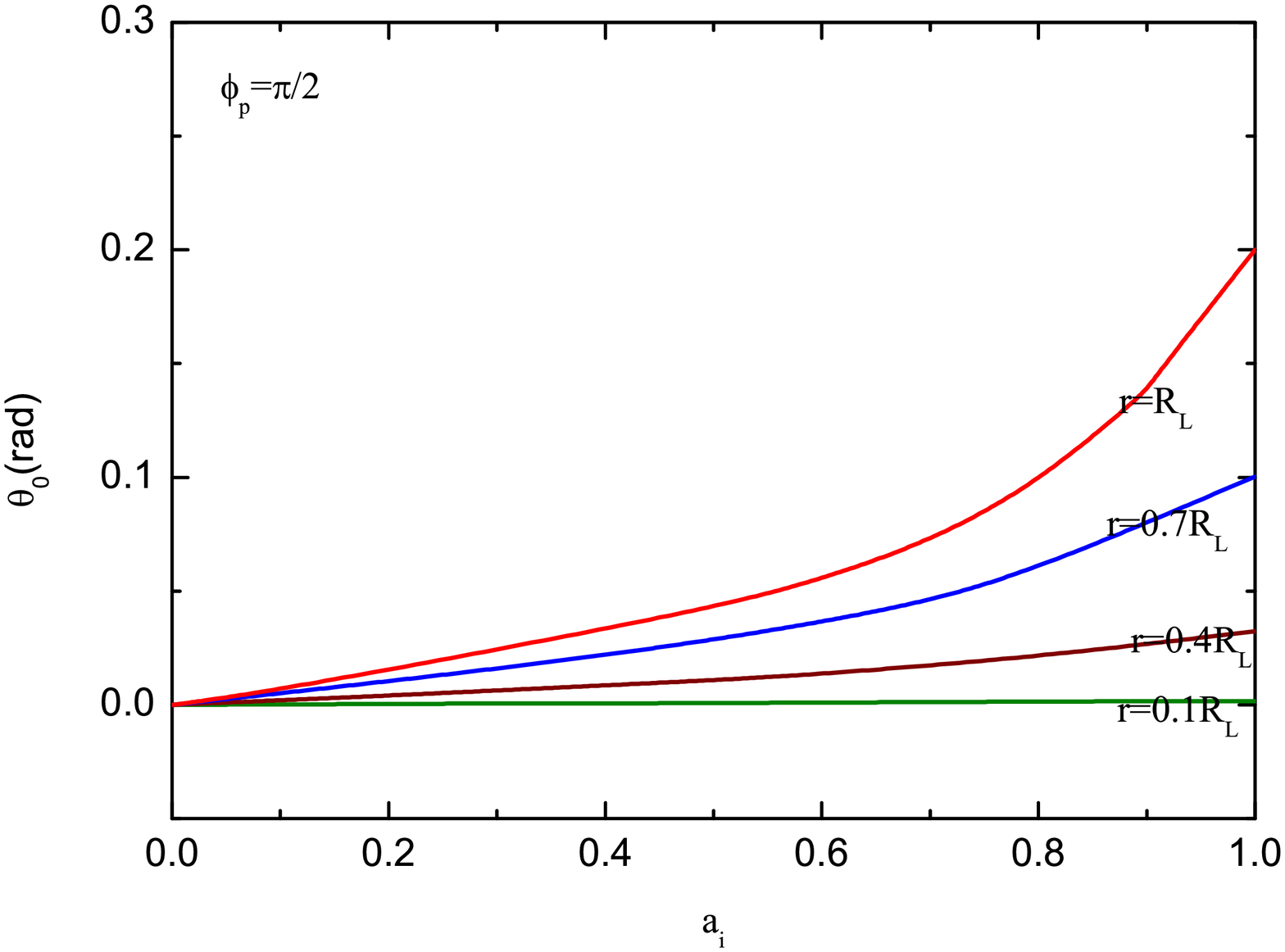}
      \caption{Changes of particle trajectories of the particle with the radial radius (left) and the zenith angle (right). Left panel: Variations of pitch angles with the radial radius $r$ for different oblique rotators along last closed field lines. Inset in the figure shows the results of the logarithm of the Y-axis. Right panel: Variations of pitch angle with $a_i$ at different positions for a $45^{\circ}$ rotator. Notes that all consequences are calculated by $\phi_{\rm p}=\frac{\pi}{2}$. }
      \label{fig:3}
\end{figure*}

We now describe our calculation procedure as follows.
\begin{itemize}
\item[(i)] The determination of a magneto-spherical structure. To calculate the particle trajectory in the open field line regions of the magnetosphere, the footpoints of the open magnetic field lines are required to be determined at first. In the magnetosphere considered here, the footpoints of the open magnetic field lines are given by,
\begin{equation}\label{Eq21}
  (x_{0,i}, y_{0,i}, z_{0,i})=\left[a_iR_{\rm p}\cos(\phi_{\rm p}), a_iR_{\rm p}\sin(\phi_{\rm p}), (R_*^2-a_i^2R_{\rm p}^2)^{\frac{1}{2}} \right] \;,
\end{equation}
where $R_{\rm P}=R_*(R_*/R_{\rm L})^{\frac{1}{2}}$ is the radius of the polar caps, $a_i$ is the scaling factor which represents the angle between the magnetic axis and the footpoint of the $i$th magnetic field line in open field line region, it ranges from $0$ (magnetic axis) to $1$ ($a_0=1$ for the last open field lines, otherwise $0 \le a_i<1$ with $i>0$), and $\phi_{\rm p}$ is the azimuthal angle on the X-Y plane corresponding to the footpoints of the field lines about the magnetic axis \citep{Zhang2009,Chang2015}. For a given $\phi_{\rm p}$, the footpoint of an open field line is determined, and then the corresponding magnetic field line is calculated by using Eq.(\ref{Eq1}) in the magnetosphere. Once a footpoint is given for a given inclination angle $\alpha$, corresponding a magnetic field line is calculated by using the Runge-Kutta method.
\item[(ii)] The estimation of the particle trajectory. A particle on polar caps of the stellar surface moves initially along a given magnetic field line, and then is accelerated rapidly by the local accelerating electric field. Its velocity $\bm{\beta}$ is given by Eq. (\ref{Eq10}). Since the direction of $\bm{\beta}$ on the influence of the local accelerating electric field and radiation reaction can deviate from the direction of the magnetic field line, forming an angle between them given by Eq. (\ref{Eq13}). Therefore, the particle trajectories at the open field-line region can be determined.
\item[(iii)] The calculation of synchro-curvature spectrum. Once the particle's trajectory is determined and the value of $\bm{\beta}$ is given, the particle's Lorentz factor is calculated by Eq.({\ref{Eq12}}). The curvature radius of the particle trajectory can be estimated by using Eq.(\ref{EqB5}). The direction, $\bm{n}$ (or $\bm{\beta}$), of the particle motion is assumed to be the emission direction of the photons and can be calculated in the inertial observer's frame. This emission direction is usually expressed by using the viewing angle $\zeta$ and rotation phase $\Phi$, which can be calculated by using $\cos\zeta=\bm{n}_z$ and $\Phi=-\Phi_n-\bm{r}\cdot \bm{n}/R_{\rm L}$, respectively, where $\Phi_n=\arctan (\bm{n}_y/\bm{n}_x)$ is the azimuthal angle. Once the viewing angle $\zeta$ is determined, the observed photons are then from the region of $\zeta-\theta_0 \leq \zeta \leq \zeta+\theta_0$. Therefore, the synchro-curvature spectrum emitted by a particle with a single Lorentz factor $\gamma_{\rm e}$ can be calculated by Eq.(\ref{Eq15}).
\end{itemize}

\begin{figure}
      \centering
      \includegraphics[width=0.5\textwidth]{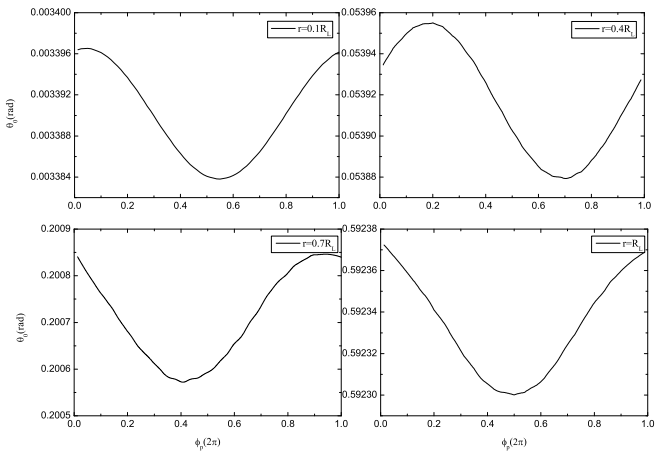}
      \caption{Changes of pitch angle with $\phi_{\rm p}$ at different radial positions for a $45^{\circ}$ rotator. From top to bottom we show the results of $r\in[0.1, 0.4, 0.7, 1]R_{\rm L}$, in which the particle moves along the last open field lines. }
      \label{fig:4}
\end{figure}

\section{Results}\label{sec:Results}

\subsection{Comparison of particle trajectories and magnetic field lines}

A particle is accelerated rapidly to ultra-relativistic velocity, and then moves outwards at the footpoint of an open magnetic field line with a given $\phi_{\rm p}$.  The trajectory of the particle with velocity $\bm{\beta}$ in a given position of the magnetosphere is determined by Eq.(\ref{Eq10}), and corresponding pitch angle is calculated by using Eq.(\ref{Eq13}). The trajectory of the particle is the function of ($r, a_i, \phi_p, \alpha$), namely the pitch angle $\theta_0$ is the function of ($r, a_i, \phi_p, \alpha$). We can calculate the velocity of radiating particles passing through each magnetospheric point in the computational domain. Each point on the 'trajectory' corresponds to the point on an open field line.

We firstly show the comparisons of the last closed field lines with the particle trajectories in the magnetospheres with $\alpha=0^{\circ},~30^{\circ}, ~60^{\circ}$, and $90^{\circ}$ in 3D Cartesian Coordinates in Fig.\ref{fig:1}. The black lines represent the last closed field lines and the red lines are the particle trajectories along the magnetic field lines (blue lines). It can be seen that the trajectories are not along the magnetic field lines and the magnetic field lines of an oblique rotator are more curved than that of an aligned one, which leads to the trajectory of accelerating electrons becoming curved, especially in the vicinity of the light cylinder. In all cases, the trajectories obviously deviate from the magnetic field lines in the vicinity of the light cylinder.

For a given $\alpha$, the discrepancies between particle trajectories and open magnetic field lines on X-Z and Y-Z planes are shown in Fig.\ref{fig:2}, where five values of $a_i =[1.0,~0.8,~0.6,~0.4,~0.2]$ are used to compare the discrepancies. In this figure, blue dash curves depict the usual open field lines and red solid curves correspond to particle trajectories. Notes that a set of curves on the same sides have the same azimuthal angle $\phi_{\rm p}$ on the polar cap. For example, a series of curves of $x>0$ for an aligned rotator are simulated through Runge-Kutta methods with $\phi_{\rm p}=0$. And then curves of $x<0$ present the situation of $\phi_{\rm p}=\pi$. The case of $\phi_{\rm p}=\pi/2$ and $3\pi/2$ are shown on the Y-Z plane. The first row of Fig.\ref{fig:2} presents the case for an aligned rotator. For an aligned rotator, the structure of the pulsar magnetosphere is symmetric about the Z-axis, which assumed that the rotation axis of the pulsar is the Z-axis. Projections of open field lines and trajectories on X-Z and Y-Z planes are the same at each $\phi_{\rm p}$. On X-Z and Y-Z planes, projections of both filed lines and trajectories are almost identical. With the increase of $\alpha$, the difference between open field lines and trajectories becomes larger. It is shown obviously for the case $\alpha=90^{\circ}$ on the Y-Z plane.

Figs. \ref{fig:3}-\ref{fig:4} show the variations of pitch angle $\theta_0$ with ($r, \theta_0, \phi_p$). The left panel of Fig.\ref{fig:3} shows the variation of the pitch angle with $r$ and $\alpha$ along the last open field line, where the variation of $\alpha$ is expressed in colorful lines. From the figure, $\theta_0$ increases with increasing $r$ and $\alpha$ when other parameters are given. Trajectories indicate that electrons initially move along the magnetic field line ($\theta_0\approx 10^{-8}-10^{-2}$), and then away from the field lines. The deviation of the trajectory from the magnetic field line relates to $\beta_{\rm D}$ and $\beta_{\rm E}$, especially the drift velocity $\beta_{\rm D}$. Because the effect of the strong magnetic field leads to relatively a small $\beta_{\rm D}$ near the neutron star, the proportion of drift velocity near the light cylinder becomes larger with decreasing magnetic field. The right panel of Fig.\ref{fig:3} shows the variation of the pitch angle with $a_i$ at different positions, which are plotted as colorful lines, where $\alpha=45^{\circ}$ is used as an example. Basically, the pitch angle increases with increasing $a_i$ (namely $\theta$), but the increase is not obvious near the neutron star (such as $r=0.1R_{\rm L}$). Fig.\ref{fig:4} shows the variation of the pitch angle with $\phi_{\rm p}$ at different positions for a $45^{\circ}$ rotator, indicating that the change of the pitch angle with $\phi_{\rm p}$ is different at the various radial position.

\subsection{The Lorentz factor and radius of curvature of the particle trajectory}

\begin{figure}
      \centering
      \includegraphics[width=0.5\textwidth]{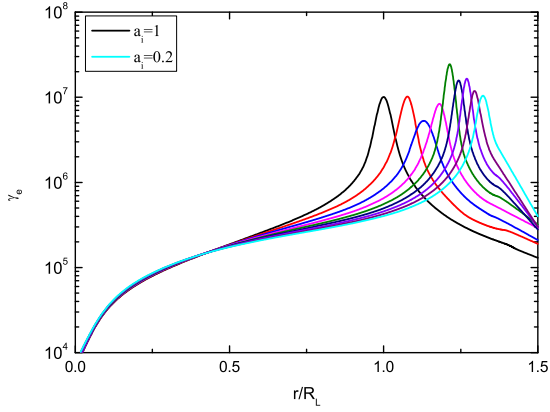}
      \caption{The Lorentz factor of the trajectories of particles. The Lorentz factor is a function of the radial distance ($r$), the $i$th open field line ($a_i$), and the azimuthal angle ($\phi_p$). The figure shows $\gamma_{\rm e}$ as a function of $r$ and $a_i$, where $a_i$ ranged from $0.2$ to $1$ is color-coded at the plane of $\phi_p=\frac{\pi}{2}$.  The Crab pulsar parameters and inclination angle $\alpha=45^{\circ}$ are used. }
      \label{fig:5}
\end{figure}

From Eq.(\ref{Eq12}), the Lorentz factor is also the function of $r, a_i, \phi_p$ and $\alpha$. Taking the inclination angle $\alpha=45^{\circ}$ as an example, we show the relationship of $\gamma_{\rm e}$ and ($r, a_i$) in the Fig.\ref{fig:5}, where the value of $a_i$ ranged from $0.2$ to $1$, is color-coded at the plane of $\phi_p=\frac{\pi}{2}$. The graphic indicates that $\gamma_{\rm e}$ is approximately $10^4-10^8$ and rises linearly in the region of inner the light cylinder.
It is worth noting that the particle trajectory obtained in this paper is based on a certain field line. The Lorentz factor of the particle does not keep increasing to infinity along the trajectory. For example, $\gamma_{\rm e}$ reaches the maximum value at the light cylinder along the last closed field line ($a_i=1$), and decreases in the region of $r>R_{\rm L}$. This is because the magnetic field lines are closed lines, when $r>R_{\rm L}$, the field lines flow to another pole, the radial distance decrease, but the particles are always moving outward. Thus, the pitch angle between the trajectory and field line becomes larger. For the open field lines with $0<a_i\le 1$, generally, the position where $\gamma_{\rm e}$ reaches its maximum value is $\eta_i R_{\rm L}$ with $\eta_i\ge 1$, $\gamma_{\rm e}$ will decrease at the position larger than $\eta_i R_{\rm L}$  (see Fig.\ref{fig:5}).

\begin{figure}
      \centering
      \includegraphics[width=0.5\textwidth]{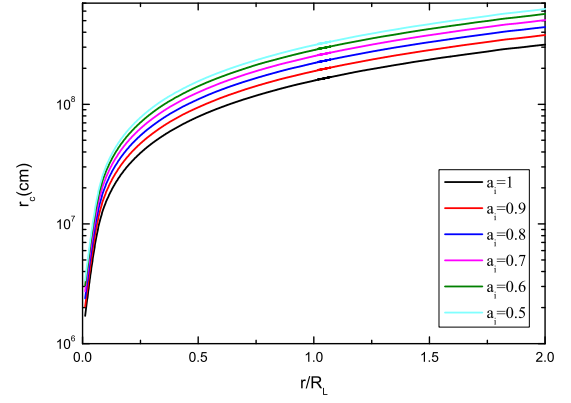}
      \caption{Variations of the radius of curvature of particle trajectories with the radial radius for an oblique rotator along the open field lines.  The parameters of $\alpha=45^{\circ}$ and $\phi_p=\frac{\pi}{2}$ are used.    }
      \label{fig:6}
\end{figure}

Fig.\ref{fig:6} shows variations of $r_{\rm c}$ with the radial radius $r$ for the case of $\alpha=45^{\circ}$ and $\phi_p=\pi/2$, the $a_i$ ranges from $1$ to $0.5$ with the interval of $0.1$. The figure indicates that $r_{\rm c}$ rises linearly from the neutron star surface to infinity, thus, the curvature radiation in the region of inner the light cylinder is extremely important.

In general, we can use Eq.(\ref{Eq15}) to calculate the synchro-curvature spectrum at any position along a particle trajectory. In Figs. \ref{fig:7} and \ref{fig:8}, we show the spectral energy distributions (SEDs) of the synchro-curvature radiation for the inclination angle with $\alpha=45^{\circ}$. The radiation region can be defined as expending from the neutron star surface to the light cylinder in the open field-line region.

\subsection{Synchro-curvature spectrum}

The SEDs emitted from different positions are shown in Fig.\ref{fig:7} with color lines. The electron Lorentz factor is calculated by using Eq. (\ref{Eq12}). Since the Lorentz factor and the pitch angle are about $10^4-10^5$ (see Fig.\ref{fig:5}) and about $10^{-8}-10^{-2}$ (Fig. \ref{fig:3}) near the surface of the neutron star, the photon energy of synchrotron radiation is very small. Therefore, the radiation in MeV - GeV band is dominated by curvature radiation. As the increase of the radial distance, the Lorentz factor $\gamma_{\rm e}$ and pitch angle $\theta_0$ increase. The cut energy of the synchrotron radiation also increases. Therefore, the contribution of synchrotron radiation near the light cylinder can not be ignored compared with previous studies.

Fig. \ref{fig:8} shows the phase-averaged spectrum along the last closed field lines with the parameters of $\alpha=45^{\circ}$ and $\zeta=55^{\circ}$ and $75^{\circ}$. Each solid curve plotted in the figure represents the case for different viewing angles. The comparison indicates that the SEDs of the case of $\zeta=75^{\circ}$ appear as an obvious double-peak structure, and the case of $\zeta=55^{\circ}$ do not. That is because that the profile of the SEDs is determined by the position where the radiation occurs, namely, it relates to the viewing angle of the observer. In summary, the inclination angle determines the geometry in the pulsar magnetosphere, and the viewing angle determines the radiation position of the observed photons, namely, the SED profile.

\section{Summary and discussion}\label{sec:summary and discussion}

In this paper, the trajectory of accelerating electrons are calculated inside the light cylinder of the pulsar magnetosphere with the \cite{Deutsch1955} electromagnetic field, where the accelerating electric field can be estimated as $E_{\|}=\frac{\textbf{E}\cdot \textbf{B}}{|B|}$ everywhere. When the structure of the electromagnetic field is given, the electron velocity can be calculated by Eq.(\ref{Eq10}), which depends on \textbf{$E$} and \textbf{$B$}. The results show that the real particle trajectory is not along the open magnetic field line but has a pitch angle between the magnetic field lines and trajectories after the effect of the radiation reaction force is included. In this case, the pitch angle is the increasing function of $r$, $a_i$ (or $\theta$), and $\alpha$.
Since the velocity consists of three components, especially for $\bm{\beta}_{\rm D}=(\bm{E}\times\bm{B})/B^2$, it causes the trajectory deviates from the magnetic field line. Trajectories of electrons along the magnetic field lines for aligned and oblique rotators are modeled and shown in Figs.\ref{fig:1} and \ref{fig:2}.

The curvature radius can be estimated corresponding to the trajectory of the particle by using Eq.(\ref{EqB2}).  In addition, we can estimate the optical depth for photons emitted at a given position as $\tau =n_{\rm ph} \sigma l(r)$, where $n_{\rm ph}$ represent the density of photon, $\sigma$ is the cross-section, and $l(r)$ is the distance from the emission location to the observer. Assuming that the density of photon (and cross-section) is the same everywhere, then the value of the optical depth mainly depends on the $l(r)$. Obviously, the distance from the neutron star surface to the observer ($l(r_{\rm NS})$) is larger than that from the light cylinder to the observer ($l(r_{\rm LC})$). The ratio of the observed  photon flux is $e^{-\tau_{\rm NS}}/e^{-\tau_{\rm LC}}<1$, namely, the photon emitted near the light cylinder is easier to capture by the observers than that emit near the neutron star surface.

\begin{figure}
      \centering
      \includegraphics[width=0.5\textwidth,angle=0]{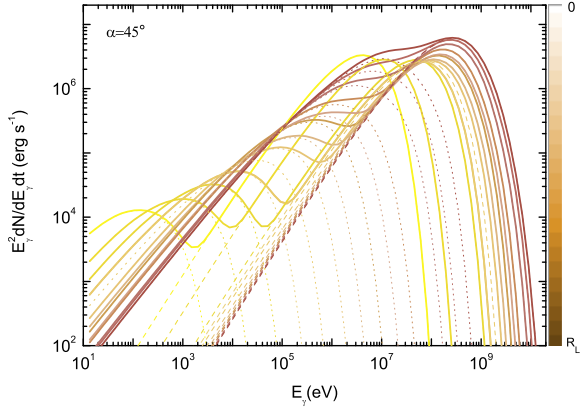}
      \caption{ The synchro-curvature spectrum of a single electron emitted along the particle trajectory. Each curve represents the SED of a given portion of the trajectory of particles along the last open field line from the neutron star surface to the light cylinder, where the position is color-coded. Note that $\alpha=45^{\circ}$ is used and the viewing angle is not considered here.}
      \label{fig:7}
\end{figure}

\begin{figure}
      \centering
      \includegraphics[width=0.5\textwidth,angle=0]{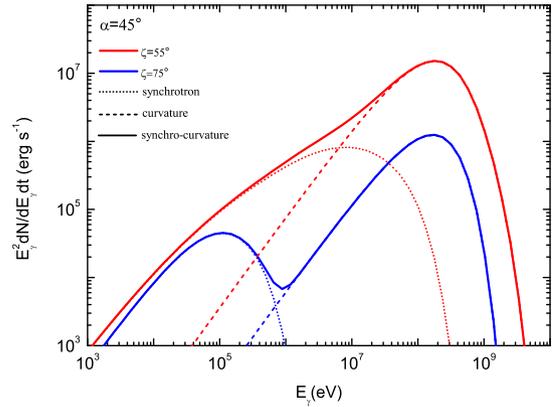}
      \caption{ Phase-averaged spectrum of a single electron with a mono-energetic form. The model parameters are: the inclination angle $\alpha=45^{\circ}$, the viewing angle $\zeta=55^{\circ}$ (red lines) and $75^{\circ}$ (blue lines), and $a_i=0.99$. Dot lines represent the spectrum of the synchrotron radiation, dash lines represent the spectrum of the curvature radiation, and solid lines represent the sum of them.}
      \label{fig:8}
\end{figure}

The emission of non-thermal photons is described through the synchro-curvature radiation of ultra-relativistic electrons. Therefore, the synchro-curvature SEDs produced by electrons are also be modeled along the electron trajectories. The calculated SEDs are shown in Figs.\ref{fig:7} and \ref{fig:8}.  Our results show that the synchro-curvature SEDs appear a double peak structure, the synchrotron radiation plays an important role in the X-ray band and curvature radiation mainly works in the GeV band, depending on the radiation position. In this paper, we can model the SEDs with only two free parameters: $\alpha$ and $\zeta$. In fact, the synchro-curvature SED with solving the particle trajectory has been obtained by using different methods \citep[e.g.,][]{Kelner2015,Torres2018}.

Finally, it should be pointed out that the particle trajectory approach has been used to describe the particle trajectory in the magnetosphere simulations \cite[e.g.,][]{KHK14,KHKB17,KBTHK18}. Recently, based on analytical approximation of vacuum electromagnetic field of a rotating dipole, polar cap plus slot gap \citep{GP2021} and  polar cap plus striped wind current sheet \citep{PM2021} have been applied to explain the observed properties in radio and gamma-ray bands. We will apply our current work to study the observed emissions from young gamma-ray pulsars in our next work.

\section{acknowledgments}

We thank the anonymous referee for his/her very constructive comments. The work of S. Chang is partially supported by the National Natural Science Foundation of China 12103046, and the Foundations of Yunnan Province 202101AU070036. The work of L. Zhang is partially supported by the National Key R \& D Program of China under Grant No. 2018YFA0404204, and the National Natural Science Foundation of China U1738211. The work of Z. J. Jiang is partially supported by the National Natural Science Foundation of China U1931113.

\section{DATA AVAILABILITY}

The data underlying this paper will be shared on reasonable request to the corresponding author.


\onecolumn
\appendix

\section{MOTION OF A CHARGED PARTICLE IN ELECTRIC AND MAGNETIC FIELDS}
\label{appendix A}

The motion of an ultra-relativistic charged particle in the electric and magnetic fields can finally be written as Eq.(\ref{Eq8}),
\begin{equation}\label{EqA1}
  (\textbf{F}_{\rm L}-\textbf{F}_{\rm rr})-\bm{\beta}[\bm{\beta}\cdot(\textbf{F}_{\rm L}-\textbf{F}_{\rm rr})]- m_{\rm e}c\gamma_{\rm e} \frac{d\bm{\beta}}{dt}=0 \;,
\end{equation}
where the Lorentz force is $\textbf{F}_{\rm L}=e(\textbf{E}+\bm{\beta}\times\textbf{B})$, and the radiation reaction force takes the first two terms of Eq.(\ref{Eq9}),
\begin{align}\label{EqA2}
  \nonumber \textbf{F}_{\rm rr} &=\frac{2e^4}{3m_{e}^{2}c^4}[\textbf{E}\times\textbf{B}+\textbf{B}\times(\textbf{B}\times\bm{\beta})+\textbf{E}(\bm{\beta}\cdot\textbf{E})] \\
                                &-\frac{2e^4\gamma_e^2}{3m_e^2c^4}\bm{\beta}[(\textbf{E}+\bm{\beta}\times\textbf{B})^2-(\textbf{E}\cdot\bm{\beta})^2] \;,
\end{align}
Here $C_1=2e^3/3m_{e}^{2}c^4$ and the dimension of $\frac{1}{C_1}$ is same as $\textbf{E}$ and $\textbf{B}$. The second term $\bm{\beta}\cdot(\textbf{F}_{\rm L}-\textbf{F}_{\rm rr})$ of Eq.(\ref{EqA1}) is a scalar, which can be defined as the coefficient $C_2$, then Eq.(\ref{EqA1}) can be written as,
\begin{equation}\label{EqA3}
   \textbf{E}+\bm{\beta}\times\textbf{B}-C_1[\textbf{E}\times\textbf{B}+\textbf{B}\times(\textbf{B}\times\bm{\beta})+\textbf{E}(\bm{\beta}\cdot\textbf{E})]-C_1\gamma_e^2\bm{\beta}[(\textbf{E}+\bm{\beta}\times\textbf{B})^2
  -(\textbf{E}\cdot\bm{\beta})^2]-C_2\bm{\beta} -\frac{m_{\rm e}c\gamma_{\rm e}}{e} \frac{d\bm{\beta}}{dt}=0 \;,
\end{equation}
where the term $[(\textbf{E}+\bm{\beta}\times\textbf{B})^2-(\textbf{E}\cdot\bm{\beta})^2]$ is also a scalar, which is defined as $C_3$, then both side of the Eq.(\ref{EqA3}) $\times \textbf{B}$ is
\begin{equation}\label{EqA4}
  \textbf{E}\times\textbf{B}+\bm{\beta}\times\textbf{B}\times\textbf{B}-C_1[\textbf{E}\times\textbf{B}\times\textbf{B}+\textbf{B}\times(\textbf{B}\times\bm{\beta})\times\textbf{B}
  +(\textbf{E}\times\textbf{B})(\bm{\beta}\cdot\textbf{E})]-C_1C_3\gamma_e^2(\bm{\beta}\times\textbf{B})-C_2(\bm{\beta}\times\textbf{B}) -\frac{m_{\rm e}c\gamma_{\rm e}}{e} (\frac{d\bm{\beta}}{dt}\times \textbf{B})=0  \;,
\end{equation}
where the velocity of the ultra-relativistic charged particle is very close to the speed of light ($v\approx c$), then $\frac{d\beta}{dt}=0$. Thus, we can obtain that $(\frac{d\bm{\beta}}{dt}\times \textbf{B})=\frac{d\beta}{dt}B\sin<\frac{d\bm{\beta}}{dt}, \textbf{B}> \bm{e_{\perp}}=0$, where $\bm{e_{\perp}}$ is the direction vector perpendicular to the $\frac{d\bm{\beta}}{dt}$ and $\textbf{B}$, as well as $<\frac{d\bm{\beta}}{dt}, \textbf{B}>$ is the included angle between vector $\frac{d\bm{\beta}}{dt}$ and $\textbf{B}$. Since $\bm{\beta}\times\textbf{B}\times\textbf{B}=(\bm{\beta}\cdot\textbf{B})\textbf{B}-B^2\bm{\beta}$ and $\textbf{B}\times(\textbf{B}\times\bm{\beta})\times\textbf{B}=\rm B^2(\textbf{B}\times\bm{\beta})$ , then
\begin{equation}\label{EqA5}
  \rm \textbf{E}\times\textbf{B}+(\bm{\beta}\cdot\textbf{B})\textbf{B}-B^2\bm{\beta}-C_1[(\textbf{E}\cdot\textbf{B})\textbf{B}-B^2\textbf{E}+B^2(\textbf{B}\times\bm{\beta})+(\textbf{E}\times\textbf{B})(\bm{\beta}\cdot\textbf{E})]
  -(C_1C_3\gamma_e^2-C_2)(\bm{\beta}\times\textbf{B})=0  \;,
\end{equation}
Eq.(\ref{EqA5}) $\cdot\bm{\beta}$, which gives
\begin{equation}\label{EqA6}
  \rm (\textbf{E}\times\textbf{B})\cdot\bm{\beta}+(\bm{\beta}\cdot\textbf{B})\textbf{B}\cdot\bm{\beta}-B^2\bm{\beta}\cdot\bm{\beta}-C_1[(\textbf{E}\cdot\textbf{B})\textbf{B}\cdot\bm{\beta}
  -B^2(\textbf{E}\cdot\bm{\beta})+B^2(\textbf{B}\times\bm{\beta})\cdot\bm{\beta}+(\bm{\beta}\cdot\textbf{E})(\textbf{E}\times\textbf{B})\cdot\bm{\beta}]
  -(C_1C_3\gamma_e^2-C_2)(\bm{\beta}\times\textbf{B})\cdot\bm{\beta}=0  \;,
\end{equation}
because of $\bm{\beta}\bot(\bm{\beta}\times\textbf{B})$, $(\bm{\beta}\times\textbf{B})\cdot\bm{\beta}=0$, the above equation can be written as,
\begin{equation}\label{EqA7}
  \rm [(\textbf{E}\times\textbf{B})+(\bm{\beta}\cdot\textbf{B})\textbf{B}-B^2\bm{\beta}-C_1(\textbf{E}\cdot\textbf{B})\textbf{B}+C_1B^2\textbf{E}-C_1(\bm{\beta}\cdot\textbf{E})(\textbf{E}\times\textbf{B})]\cdot\bm{\beta}=0
  \;,
\end{equation}
there are two situations for solving Eq.(\ref{EqA7}): (1) $\bm{\beta}\neq 0$ and the term of brackets is equal to 0,  and (2) the term in brackets is perpendicular to $\bm{\beta}$. It is so complex to implement for situation (2), thus only the situation (1) is considered in this paper. Therefore, the form of the relative velocity of relativistic particle can be obtained as
\begin{equation}\label{EqA8}
  \bm{\beta}=\frac{[1-C_1(\bm{\beta}\cdot\textbf{E})](\textbf{E}\times\textbf{B})+[(\bm{\beta}\cdot\textbf{B})-C_1(\textbf{E}\cdot\textbf{B})]\textbf{B}+C_1B^2\textbf{E}}{B^2}  \;.
\end{equation}

Eq.(\ref{EqA8}) also can be expressed as $\bm{\beta}=\bm{\beta}_{\rm D}+\bm{\beta}_{\rm B}+\bm{\beta}_{\rm E}$, where $\bm{\beta}_{\rm D}$ represent the drift velocity (the first term of Eq.(\ref{EqA8})), $\bm{\beta}_{\rm B}$ represent the component of velocity along the magnetic field lines (the second term of Eq.(\ref{EqA8})), and $\bm{\beta}_{\rm E}$ represent the component of velocity along the direction of electric field (the third term of Eq.(\ref{EqA8})). From Eq.(\ref{EqA8}), we can know that $\bm{\beta}_{\rm E}=C_1B^2\textbf{E}/B^2$ and $|\bm{\beta}\cdot\textbf{E}|=|\bm{\beta}_{\rm E}|\cdot |E|$, then $C_1|\bm{\beta}\cdot\textbf{E}|=C_1^2|E|^2\leq 10^{-12}$, where $E_{\rm max}\approx 10^{10}$ (V/m) and $C_1=2e^3/3m_e^2c^4\approx 1.1\times 10^{-16}$ (1/G) . Consequently, $1-C_1(\bm{\beta}\cdot\textbf{E})$ is approximately 1 and the component of the drift velocity can be estimated as $\bm{\beta}_{D}=(\textbf{E}\times\textbf{B})/B^2$.

In this paper, it is defined that $B_0=(\bm{\beta}\cdot\textbf{B})-C_1(\textbf{E}\cdot\textbf{B})$ and $E_0=C_1B^2$ . $B_0$ and $E_0$ represent the quantities which have the same dimensions as $\textbf{E}$ and $\textbf{B}$. Thus, the velocity of electrons is written as,
\begin{equation}\label{EqA9}
  \bm{\beta}=\frac{(\textbf{E}\times\textbf{B})+\rm B_0\textbf{B}+E_0\textbf{E}}{\rm B^2}  \;,
\end{equation}
Similarly, we can also the velocity of positrons,
\begin{equation}\label{EqA11}
  \bm{\beta}=\frac{(\textbf{E}\times\textbf{B})-(\rm B_0\textbf{B}+E_0\textbf{E})}{\rm B^2}  \;,
\end{equation}
where $B_0$ can be estimated by $\beta^2=1$,
\begin{align}\label{EqA10}
    \nonumber  &  \rm E_0=C_1B^2  \;, \\
               &  \rm B_0=\rm{sign}(\bm{\beta}\cdot\textbf{B}) \sqrt{\left(\frac{- E_0(\textbf{E}\cdot\textbf{B})\pm B \sqrt{B^4-(\textbf{E}\times\textbf{B})^2}}{B^2}\right)^2}  \;.
\end{align}

\section{Calculation of the radius of curvature } \label{appendix B}

The motion of an ultrarelativistic charged particle in the electric ($\bm E$) and magnetic ($\bm B$) fields is described by the system of equations\citep{Landau&Lifshitz1987},
\begin{align}\label{EqB1}
  \nonumber \frac{d\bm{r}}{dt}         & = c\bm{\beta} \\
            \frac{d\gamma_{\rm e}}{dt} & = \frac{e}{m_{\rm e}c}\left( \bm{\beta}\cdot\bm{E} \right)- \frac{2e^2}{3m_{\rm e}c}\frac{\gamma_{\rm e}^4}{r_{\rm c}^2} \;,
\end{align}
where the curvature radius of the trajectory is
\begin{equation}\label{EqB2}
  \frac{1}{r_{\rm c}}=|K|=|(\bm{\beta}\cdot \nabla)\bm{\beta}| \;,
\end{equation}
the velocity vector of the trajectory $\bm{\beta}$ and Hamiltonian $\nabla$  can be described as $\bm{\beta}=\beta_x \bm{i}+\beta_y \bm{j}+\beta_z \bm{k}$ and $\nabla=\bm{i}\frac{\partial}{\partial x}+\bm{j}\frac{\partial}{\partial y}+\bm{k}\frac{\partial}{\partial z}$ in the Cartesian coordinates. We can get
\begin{equation}\label{EqB3}
 \begin{split}
       \bm{\beta}\cdot \nabla & =\left(\beta_x \bm{i}+\beta_y \bm{j}+\beta_z \bm{k}\right)\cdot \left(\bm{i}\frac{\partial}{\partial x}+\bm{j}\frac{\partial}{\partial y}+\bm{k}\frac{\partial}{\partial z}\right)   \\
                              & =\left(\frac{\partial \beta_x}{\partial x}+\frac{\partial \beta_y}{\partial y}+\frac{\partial \beta_z}{\partial z}\right)  \\
                              & = |\frac{d\bm{\beta}}{d\bm{r}}| \;,
 \end{split}
\end{equation}
From Eq.(\ref{EqB1}) and (\ref{EqB3}), we can easily obtain $\left(\bm{\beta}\cdot \nabla\right)\bm{\beta}  =|\frac{d\bm{\beta}}{d\bm{r}}|\frac{d\bm{r}}{c dt}$, then the curvature radius of the trajectory can be calculated as,
\begin{equation}\label{EqB4}
  \frac{1}{r_{\rm c}}=|K|=|(\bm{\beta}\cdot \nabla)\bm{\beta}| =\frac{1}{c}|\frac{d\bm{\beta}}{dt}| \;.
\end{equation}

As we know the expression of $d\bm{\beta}/dt$ from Eq.(\ref{Eq7}), Eq.(\ref{EqB4}) can be written as,
\begin{equation}\label{EqB5}
  {r_{\rm c}} =\frac{m_{\rm e}c^2\gamma_{\rm e}}{\sqrt{\left\{ (\textbf{F}_{\rm L}-\textbf{F}_{\rm rr})-\bm{\beta}[\bm{\beta}\cdot(\textbf{F}_{\rm L}-\textbf{F}_{\rm rr})] \right\}^2}} \;.
\end{equation}
Note that $\bm{F_{\rm L}}$ and $\bm{F_{\rm rr}}$ can be estimated by Eq.(\ref{EqA2}).


\bsp	
\label{lastpage}

\begin{thebibliography}{}
\bibitem[Abdo et al.(2009)]{Adbo2009}
Abdo, A. A., et al. Science, 2009, 325: 840-844.

\bibitem[Abdo et al. (2010)]{Adbo2010}
Abdo, A. A., et al. ApJs, 2010 187: 460-494.

\bibitem[Abdo et al.(2013)]{Adbo2013}
Abdo, A. A., et al. 2013, ApJs, 208, 17(59pp).

\bibitem[Abdalla, H., et al.(2018)]{HESS2018}
Abdalla, H., et al.,(H.E.S.S. Collaboration), 2018, A\&A, 620, A66.

\bibitem[Ajello et al.(2017)]{Ajello2017}
Ajello, A. A., et al. 2017, ApJs, 232, 18.

\bibitem[Arons (1983)]{Arons83}
Arons, J. 1983, ApJ, 266, 215

\bibitem[Bai \& Spitkovsky (2010)]{Bai2010}
Bai X. -N., Spitkovsky A., 2010, ApJ, 715, 1270

\bibitem[Arons (2015)]{BKHK2015}
Brambilla, G., Kalapotharakos, C., Harding, A. K., \& Kazanas, D. 2015, ApJ, 804, 84

\bibitem[Cerutti, Philippov \& Spitkovsky(2016)]{CPS2016}
Cerutti, B., Philippov A., Spitkovsky A., 2016, MNRAS, 457, 2401.

\bibitem[Chang et al.(2015)]{Chang2015}
Chang, S., Zhang, L., \& Li, X., 2015, RAA, Vol.15, No.12, 2229.

\bibitem[Chang et al.(2018)]{Chang2018}
Chang, S., Zhang, L., Li, X., \& Jiang, Z. J., 2018, MNRAS, 475, 2185.

\bibitem[Chang \& Zhang(2019a)]{Chang2019a}
Chang S., \& Zhang L., 2019a, MNRAS, 483, 1796.

\bibitem[Chang et al.(2019b)]{Chang2019b}
Chang, S., Zhang, L., Li, X., \& Jiang, Z. J., 2019b, MNRAS, 488, 4288.

\bibitem[Cheng et al. (1986)]{Cheng1986}
Cheng, K. S., Ho, C., \& Ruderman, M. ApJ, 1986, 300, 500.

\bibitem[Cheng \& Zhang (1996)]{Cheng&Zhang1996}
Cheng K. S., \& Zhang J. L., 1996, ApJ, 463, 271.

\bibitem[Cheng et al. (2000)]{CRZ2000}
Cheng, K. S., Ruderman, M. \& Zhang, L. ApJ, 2000, 537, 964.

\bibitem[Contopoulos et al. (2016)]{Contopoulos2016}
Contopoulos, I. 2016, JPlPh, 82, 6303

\bibitem[Contopoulos et al. (2020)]{CPS2020}
Contopoulos, I., P\'{e}tri, J., Stefanou, P. 2020, MNRAS, 491, 5579


\bibitem[Coroniti (1990)]{Coroniti1990}
Coroniti, F. V. 1990, ApJ, 349, 538.


\bibitem[Daugherty \& Harding (1994)]{DH94}
Daugherty, J. K. \& Harding, A. K. 1994, ApJ, 429, 325

\bibitem[Daugherty \& Harding (1996)]{DH96}
Daugherty, J. K. \& Harding, A. K. 1996, ApJ, 458, 278

\bibitem[Deutsch (1955)]{Deutsch1955}
Deutsch, A. 1955, Ann. d’Astrophys., 18.

\bibitem[Du et al. (2011)]{Du2011}
Du, Y. J., Han, J. L., Qiao, G. J., Chou, C. K. 2011, ApJ, 731, 2

\bibitem[Du et al. (2012)]{Du2012}
Du, Y. J., Qiao, G. J., Wang, W. 2012, ApJ, 748, 84

\bibitem[Dyks \& Rudak(2003)]{Dyks2003}
Dyks J., Rudak B., 2003, ApJ, 598, 1201.


\bibitem[Fang \& Zhang (2010)]{Fang2010}
Fang J., \& Zhang L., 2010, ApJ, 709, 605.



\bibitem[Giraud \& P\'{e}tri  (2020)]{GP2020}
Giraud, Q., \&  P\'{e}tri, J. 2021, A\&A, 639, A75

\bibitem[Giraud \& P\'{e}tri  (2021)]{GP2021}
Giraud, Q., \&  P\'{e}tri, J. 2021, A\&A, 654, A86

\bibitem[Harding et al. (2008)]{Harding2008}
Harding, A. K., Stern, J. V., Dyks, J., Frackowiak, M.

\bibitem[Kalapotharakos et al. (2012)]{KHKC12}
Kalapotharakos, C., Harding, A. K. Kazanas, D., \& Contopoulos, I. 2012, ApJL, 754, L1.


\bibitem[Kalapotharakos et al. (2014)]{KHK14}
Kalapotharakos, C., Harding, A. K. \& Kazanas, D., 2014, ApJ, 793, 97.

\bibitem[Kalapotharakos et al. (2017)]{KHKB17}
Kalapotharakos, C., Harding, A. K., Kazanas, D. \& Brambilla, G. 2017, ApJ, 842, 80.

\bibitem[Kalapotharakos et al. (2018)]{KBTHK18}
Kalapotharakos, C., Brambilla, G., Timokhin, A., Harding, A. K., \& Kazanas, D., 2018, ApJ, 857, 44.

\bibitem[Kelner et al. (2015)]{Kelner2015}
Kelner, S. R., Prosekin, A. Yu. \& Aharonian, F. A., 2015, ApJ, 149, 33.

\bibitem[Kuiper \& Hermsen(2015)]{Kuiper2015}
Kuiper, L., \& Hermsen, W., 2015, MNRAS, 449, 3827.

\bibitem[Landau \& Lifshitz(1987)]{Landau&Lifshitz1987}
Landau, L., \& Lifshitz, E., 1987, The Classical Theory of Fields (Oxford:Butterworth-Heinemann).

\bibitem[Michel \& Li(1999)]{Michel&Li1999}
Michel, F. C., \& Li, H., Physics Reports, 1999, 318, 227-297.

\bibitem[Muslimov \& Harding (2003)]{MH03}
Muslimov, A. G., \& Harding, A. K. 2003, ApJ, 588, 430

\bibitem[Muslimov \& Harding (2004)]{MH04}
Muslimov, A. G., \& Harding, A. K. 2004, ApJ, 606, 1143

\bibitem[P\'{e}tri \& Kirk (2005)]{PK2005}
P\'{e}tri, J. \& Kirk, J. G. 2005, ApJ, 627, L37

\bibitem[P\'{e}tri  (2009)]{Petri2009}
P\'{e}tri, J. 2009, A\&A, 503, 13

\bibitem[P\'{e}tri\& Mitra (2021)]{PM2021}
 P\'{e}tri, J. \& Mitra, D., 2021, A\&A, 654, A106

\bibitem[Philippov et al. (2015)]{PSC15}
Philippov, A. A., Spitkovsky, A., Cerutti, B. 2015, ApJL, 801, L19

\bibitem[Qiao et al. (2004)]{Qiao2004}
Qiao, G. J., Lee, K. J., Wang, H. G., Xu, R. X., Han, J. L. 2004, ApJL, 606, L49

\bibitem[Ruderman \& Sutherland (1975)]{RS75}
Ruderman, M. A. \& Sutherland, P. G. 1975, ApJ, 196, 51

\bibitem[Torrres (2018)]{Torres2018}
Torres, D. F., 2018, Nature Astronomy, 2, 247.

\bibitem[Vigan\`{o} \& Torres (2019)]{Vigano2019}
Vigan\`{o}, D., \& Torres, D. F., 2019, MNRAS, 490, 1437.

\bibitem[Zhang \& Yuan(1998)]{ZhangYuan1998}
Zhang, J. L. \& Yuan, Y. F., 1998, ApJ, 487, 370.

\bibitem[Zhang \& Cheng (1997)]{Zhang1997}
Zhang, L., \& Cheng, K. S., 1997, ApJ, 487, 370.

\bibitem[Zhang \& Cheng(2003)]{Zhang2003}
Zhang L., Cheng K. S., 2003, A\&A, 398,639.

\bibitem[Zhang, Fang \& Chen(2007)]{Zhang2007}
Zhang L., Fang J., Chen S. B., 2007, ApJ, 666, 1165.

\bibitem[Zhang \& Li (2009)]{Zhang2009}
Zhang, L., \& Li, X. 2009, ApJ, 707, L169.

\end{thebibliography}
\end{document}